\documentclass[fleqn,usenatbib]{mnras}

\usepackage{newtxtext,newtxmath}

\usepackage[T1]{fontenc}
\usepackage{ae,aecompl}

\usepackage{graphicx}	
\usepackage{amsmath}	
\usepackage{amssymb}	
\usepackage{color}      
\usepackage{float}

\newcommand{\kms}{\text{km} \ \text{s}^{-1}}
\newcommand{\SMpc}{M_{\sun} \ \text{pc}^{-2}}
\newcommand{\Ropt}{R_{\rm opt}}
\newcommand{\Vres}{V_{\rm res}}
\newcommand{\Vsm}{V_{\rm sm}}
\newcommand{\MHI}{M_\text{\HI}}
\newcommand{\Sint}{S_{\rm int}}
\newcommand{\rms}{\sigma_{\rm RMS}}
\newcommand{\rmsin}{\sigma_{\rm input}}
\newcommand{\wtw}{\omega_{20}}
\newcommand{\wft}{\omega_{50}}
\newcommand{\SN}{S/N}
\newcommand{\SNPN}{S/N_{\rm peak}}
\newcommand{\Afr}{A_{\rm fr}}
\newcommand{\HI}{H{\sc i}}

\newcommand{\dfg}{\Delta f_\text{gas}}
\newcommand{\cp}{\citep}
\newcommand{\ct}{\citet}

\title[xGASS: Robust global \HI\ asymmetries]{xGASS: Robust quantification of asymmetries in global \HI\ spectra and their relationship to environmental processes}

\author[A. B. Watts et al.]{
Adam B. Watts,$^{1,2}$\thanks{E-mail: adam.watts@research.uwa.edu.au}
Barbara Catinella,$^{1,2}$
Luca Cortese,$^{1,2}$
and Chris Power$^{1,2}$
\\
$^{1}$International Centre for Radio Astronomy Research, The University of Western Australia, Crawley, WA, Australia\\
$^{2}$ARC Centre of Excellence for All-Sky Astrophysics in 3 Dimensions (ASTRO3D)\\
}

\date{Accepted 2020 January 10. Received 2020 January 10; in original form 2019 September 3}

\pubyear{2020}

\begin{document}
\label{firstpage}
\pagerange{\pageref{firstpage}--\pageref{lastpage}}
\maketitle

\begin{abstract}
We present an analysis of asymmetries in global \HI\ spectra from
the extended GALEX Arecibo SDSS Survey (xGASS), a stellar mass-selected and gas
fraction-limited survey which is representative of the \HI\ properties of galaxies in the
local Universe. We demonstrate that the asymmetry in a \HI\ spectrum is strongly linked to its signal-to-noise meaning that, contrary to what was done in previous works, asymmetry distributions for different samples cannot be compared at face value. We develop a method to account for noise-induced asymmetry and find that the typical galaxy  detected by xGASS exhibits higher asymmetry than what can be attributed to noise alone, with 37\% of the sample showing asymmetry greater than 10\% at an 80\% confidence level. We find that asymmetric galaxies contain, on average, 29\% less \HI\ mass compared to their symmetric counterparts matched in both stellar mass and signal-to-noise. We also present clear evidence that satellite galaxies, as a population, exhibit more asymmetric \HI\ spectra than centrals and that group central galaxies show a slightly higher rate of \HI\ asymmetries compared to isolated centrals. All these results support a scenario in which environmental processes, in particular those responsible for gas removal, are the dominant driver of asymmetry in xGASS.
\end{abstract}

\begin{keywords}
galaxies: ISM -- radio lines: galaxies -- galaxies: kinematics and dynamics -- galaxies: evolution 
\end{keywords}



\section{Introduction}
The gas reservoirs of galaxies are their fundamental building blocks,  providing the potential for future star formation  and the fuel for  nuclear accretion. Understanding their  properties and dynamics is essential to completing our picture of galaxy formation and evolution. Atomic hydrogen (\HI) has proven to be a powerful tracer of these gas reservoirs though its 21cm emission line. It is typically detectable to 2-3 times the radius of the optically bright stellar component of galaxies \cp{giovanelli88,wang14} where baryons are less gravitationally bound and more sensitive to galaxy interactions and environmental processes \cp{gunn72,kenney04,stevens17}. While to date there exists $\sim500$ spatially resolved observations of the \HI\ in galaxies in the local Universe \cp{wang16} the majority of the data are unresolved, in the form of the spatially integrated global \HI\ spectrum produced by blind surveys such as the  Arecibo Legacy Fast ALFA survey \cp[ALFALFA,][]{giovanelli05,haynes18}. As radio astronomy enters the era of the Square Kilometer Array (SKA), precursor surveys such as  the Widefield ASKAP L-band Legacy All-sky Blind Survey \cp[WALLABY,][]{koribalski09} are predicted to detect over $5 \times 10^{5}$ objects, the majority of which will be spatially unresolved, global \HI\ spectra. If we wish to exploit these future surveys to their full potential we need to extract as much information as we can from global \HI\ spectra. 

The global \HI\ emission line spectrum contains information such as: the integrated flux which gives the total \HI\ mass of a galaxy, its width the projected maximum rotational velocity and therefore the dynamical mass, and its centroid provides a precise measurement  of the cosmological redshift. These properties have been used extensively to study a number of relations such as the gas fraction scaling relations \cp[e.g.][]{brown15,catinella18}, \HI\ mass function \cp[e.g.][]{zwaan05,jones18},  baryonic Tully-Fisher relation \cp[e.g.][]{mcgaugh00,papastergis16}, and the Kennicutt-Schmidt star formation law \cp[e.g.][]{kennicutt98}. 

Observations have shown that the \HI\ in galaxies frequently shows signatures of disturbance in both their spatially resolved and global observations \cp{sancisi76, baldwin80}. Smaller scale structure in  global \HI\ profiles, such as departures from symmetry, thus indicate deviation from a uniform gas distribution and/or kinematics in the galaxy \cp{swaters99,kornreich00}. Historically, high rates of asymmetry in the global \HI\ profiles of galaxies have been observed throughout the literature. Qualitative assessment of global \HI\ asymmetry by \ct{richter94} found that at least 50\% of isolated galaxies were asymmetric, leading them to state that ``asymmetries in disc galaxies may be the rule, rather than the exception". This statement was reinforced by later quantitative studies: \ct{haynes98} found an asymmetry rate of  50\% in their sample of isolated galaxies, and \ct{matthews98} found  77\%  in their sample of isolated late-type spiral galaxies. \ct{espada11} investigated the rate of asymmetries intrinsic to disc galaxies using a sample drawn from the AMIGA  \cp[Analysis of the interstellar Medium of Isolated GAlaxies,][]{verdes05} project, selected to be some of the most isolated galaxies in the Universe. They found their asymmetry distribution to be well described by a half-Gaussian with 9\% of galaxies showing asymmetry in excess of 2$\sigma$ and only 2\% in excess of 3$\sigma$.

The observed high rate of asymmetries, especially in isolated galaxies, implies that they must be either self sustaining or frequently induced otherwise they would be washed out within a few dynamical timescales. These high rates of asymmetry have also made it difficult to determine or single out the drivers of asymmetries observationally, as there are often conflicting results between studies, or significant scatter in  correlations, supporting the presence of relationships between global \HI\ asymmetries and  optical asymmetries \cp{haynes98,kornreich00,bok19}, presence of companions \cp{wilcots04}, morphology \cp{matthews98, espada11}, and environment \cp{haynes98,espada11,scott18}. 

The lack of strong correlation between individual asymmetries and galaxy properties has led the community to focus their attention on the shape of the distribution of global \HI\ profile asymmetries. Using this approach, there is growing evidence that environment must be one of the dominant drivers of asymmetry in global \HI\ spectra. \ct{scott18} showed that 16\% and 26\% of the galaxies with \HI\ detections in the Abell 1367 and Virgo clusters have asymmetry greater than the \ct{espada11} 3$\sigma$ level. This is supported by recent work by \ct{bok19}, who showed that global \HI\ asymmetries are more frequent in close galaxy pairs compared to isolated systems using a sample of $\sim600$ ALFALFA galaxies. 
 
The picture is less clear for isolated galaxies, and a number of drivers have been proposed \cp[see the review by][]{jog09} such as the accretion of gas \cp{bournaud05,sancisi08, espada11, ramirez18}, eccentric instabilities \cp{lovelace99}, wakes or distortions in dark matter haloes \cp{jog97,weinberg98, angiras06,angiras07, vaneymeren11a, vaneymeren11b}, and repeated or recent (but optically undetected) minor mergers \cp{zaritsky97, portas11,ramirez18}.

To date there hasn't been a robust and comprehensive quantification of global \HI\ asymmetry as a function of galaxy properties. This is, in part, due to the lack of a suitable dataset. The majority of \HI\ observations come from blind surveys such as ALFALFA, which predominantly detect the most \HI\ rich objects in their volume, creating a bias toward disc-dominated, star-forming galaxies. 
In this work we investigate global \HI\ asymmetries using the xGASS sample \cp{catinella18}, a stellar mass-selected survey designed to be representative of the gas properties of galaxies over a range of galaxy properties, making it an ideal sample to constrain the relationships between galaxy properties and global \HI\ asymmetries. This paper is organised as follows: in \S\ref{sec:sample} we introduce our xGASS sample, in \S\ref{sec:methods} we describe our \HI\ spectrum fitting and asymmetry measurement, and in \S\ref{sec:xGasym} we present the xGASS asymmetry distribution. In \S\ref{sec:noisecorr} we quantify the effect of noise on asymmetry measurement, and we revisit the xGASS asymmetry rate and define an asymmetric population in  \S\ref{sec:asympop}. We investigate the \HI\ properties of the asymmetric population and compare the asymmetry distributions of galaxies in different environments in \S\ref{sec:asymenv}, we place our results in context of the literature in \S\ref{sec:discuss}, and  we conclude in \S\ref{sec:concl}. 

\section{Sample} \label{sec:sample}
The xGASS sample \cp{catinella10,catinella18} is a stellar mass-selected \HI\ survey of $\sim1200$ galaxies with  $10^{9} < M_{\star}[M_{\sun}]< 10^{11.5}$ and $0.01<z<0.05$. The sample covers the \HI\ gas fraction range $-2< \log_{10}(\MHI/M_{\star})<1$ and is representative of the \HI\ fractions of galaxies over a comprehensive range in stellar mass, stellar surface density $(\mu_{\star})$, specific star formation rate (sSFR) and NUV-r colour. Additionally galaxies are classified as satellites, group central (the most massive galaxy in their group), or isolated centrals (a galaxy with no detectable satellites) using the SDSS  DR7 Group B({\sc ii}) catalogue corrected for shredding \cp{yang07,janowiecki17}.

\begin{figure*}
    \centering
    \includegraphics[width = \textwidth]{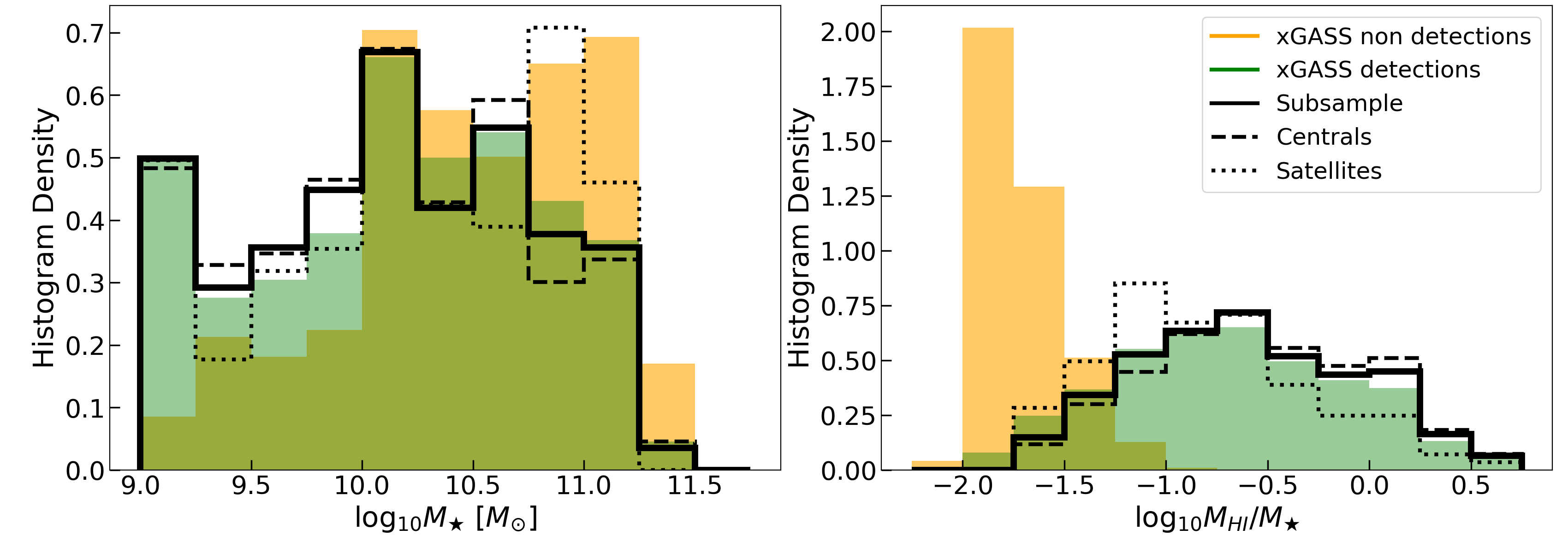}
    \caption{Sample properties: density normalised distributions of stellar mass (left) and \HI\ gas fraction (right). xGASS detections (excluding confused galaxies) are shown as a green filled histogram while the non-detections are shown as an orange filled histogram. Our selected sub-sample is shown as a solid black line, and distributions of central and satellite galaxies in the sub-sample are shown as black dashed and dotted lines respectively.}
    \label{fig:sample}
\end{figure*}

Starting with the 804 \HI\ detections (formal \& marginal) in xGASS, we excluded 108 galaxies flagged as confused as most of the observed \HI\ is physically associated to another galaxy within the beam. We remove an additional seven galaxies which have radio frequency interference (RFI) overlapping their spectrum, and a further 127 galaxies below our signal-to-noise ($\SN$) cut. Thus the final number of galaxies studied in this work is 562. Of these, 439 are classified as centrals and 113 as satellites.   In Fig. \ref{fig:sample} we show the density normalised distributions of stellar mass and gas fractions of our sub-sample compared to the xGASS parent sample, split into \HI\ detections and non-detections. As expected our sub-sample shows a close correspondence to the xGASS detections both in stellar mass and gas fraction. Compared to the overall xGASS sample the sub-sample shows a bias toward lower stellar masses as the gas fractions of these galaxies are further above the survey detection threshold. We also divide our sub-sample into satellite and central (group \& isolated) galaxies in Fig. \ref{fig:sample}. The central galaxies show the smallest  variations from the sub-sample distributions as they are the larger sample, and aside from the small excess above 10$^{10.75}\ M_{\sun}$ and deficit between 10$^{9.25}\ M_{\sun}$ and 10$^{10}\ M_{\sun}$ the satellite galaxies show a similar stellar mass distribution to the sub-sample. The gas fractions of the satellite galaxies are systematically lower than the centrals, consistent with the lower gas fractions observed in galaxies which reside in higher mass haloes  \cp{catinella13}.

\section{Methods} \label{sec:methods}

\subsection{Spectra fitting} \label{subsec:fits}
 
Consistent and robust measurement limits are required to characterise asymmetry. We fit the busy function \cp{westmeier14} to our sample of 689 \HI\ detections not affected by beam confusion or RFI as it is fast, accurate, flexible, and automatically provides an estimate of the RMS noise $(\rms)$. Each fit was visually inspected to ensure that it is an accurate parameterisation of the spectrum within the noise limits, and we force some fit parameters if necessary to prevent unphysical solutions. The fits were visually classified into three categories based on fit performance.  Out of the 689 galaxies, 582 fits were successful, 124 lacked distinct double or single peak features and were best described by a top-hat spectrum, and 89 were too irregular to be successfully fit. In the case of unsuccessful fits we forced the busy function to fit the edges of the spectrum, so that we have sensible bounds on the profile, and call these `edge fits'. 

The limits of each spectrum are defined as where the busy function fit equals 20\%  of the fits peak flux, using the individual values for each peak in double-horn spectra and the flux of the flat section in top-hat spectra. In the case of edge fits we define the limits to be where the fit equals the measured $\rms$. Using these limits we calculate the integrated flux ($\Sint$), and velocity width at the 20\% level ($\wtw$). It is worth noting that while the busy function provides a robust parameterisation of  \HI\ spectra, it may average over or ignore flux enhancements (or depressions) in a spectrum which are indistinguishable from the measurement noise. Due to this we use our fits to define the limits of each profile and measure $\wtw$ and $\rms$, but \emph{measure $\Sint$ from the observed spectrum} rather than the fit, so that we capture these variations in the observed profiles.

Although there are already published values for the integrated flux and velocity width of each spectrum in the  xGASS catalogue \cp{catinella18} we have made our own measurements of these parameters using our fits as they provide a less subjective measure of the limits of each spectrum and allow us to measure velocity widths at the 20\% level, compared to the published xGASS values measured at the 50\% level ($\wft$), to be consistent with previous global \HI\ asymmetry studies \cp[e.g.][]{espada11,scott18,bok19}. We measure a median relative difference of -4.2\% and -4.3\% in integrated flux and velocity width compared to the existing xGASS measurements, but this can be attributed to differences in the measurement techniques and identification of spectrum peaks. xGASS integrates spectra between limits defined by lines fit to the edges of each profile \cp[e.g.][]{catinella07}, whereas we integrate our spectra between the $\wtw$ velocity limits. The 1$\sigma$ relative scatter estimated from the median absolute deviation between the existing xGASS measurements and our measurements for the integrated fluxes and velocity widths are 3.4\% and 6.4\%, respectively. 

Using $\Sint$ and $\wtw$  we calculate the  $\SN$ using the ALFALFA definition \cp{saintonge07} which characterises the average flux over the profile,
\begin{equation} \label{eq:SN}
   \SN = \frac{\Sint/\wtw}{\rms}\sqrt{\frac{1}{2}\frac{\wtw}{\Vsm}},
\end{equation}
where $\Vsm$ is the smoothed velocity resolution. We adopt this definition of $\SN$ as it is representative of the ability to detect and measure a spectrum; a wider profile with the same $\Sint$ and $\rms$ will be harder to detect. Asymmetry studies typically set a lower limit of $\SN>10$ to ensure that spectra are of high enough quality to reliably recover asymmetry \cp[e.g.][]{espada11,bok19}. In this work we set a lower threshold of $\SN\geq7$ (see \S\ref{sec:asympop}). This leaves us with 562 galaxies in our final sample; 456 good fits, 81 top-hat fits and 25 edge fits.

\subsection{Asymmetry measurement} \label{subsec:Afr}
 We quantify the asymmetry of a spectrum using the integrated flux ratio parameter $\Afr$ \cp{haynes98} as it is a common parameterisation adopted in the literature \cp[e.g.][]{espada11,scott18, bok19} and analogous to flux ratio parameters presented in other studies \cp[e.g.][]{matthews98,bournaud05}. 
It is defined as 
\begin{equation} \label{eq:Afr}
    \Afr = 
        \begin{cases}
            A & A \geq 1\\
            1/A & A < 1
        \end{cases},
\end{equation}
where A is the ratio of the integrated flux on the two halves of the spectrum split by the middle velocity $V_M = 0.5(V_{\rm min} + V_{\rm max})$:
\begin{equation} \label{eq:A}
    A = \frac{
        \int_{V_{M}}^{V_{\rm max}} S_{\nu} d\text{v}
        }{
        \int_{V_{\rm min}}^{V_{M}} S_{\nu} d\text{v}
        }.
\end{equation}
We adopt $V_{\rm max}$ and $V_{\rm min}$ as the  limits defined from our fits used for measuring $\Sint$ and $\wtw$. Like $\Sint$, $\Afr$ is calculated from each observed spectrum rather than its busy function fit. 

Aside from the effects of measurement noise on integrated fluxes, which we quantify in \S\ref{subsec:SNafrcor}, there are four potential issues that could affect the measurement of $\Afr$. The first one is the effect of noise on a spectrum's peaks.  Should the value of a peak be reduced or increased the measurement limits would shift, resulting in a shift of the middle velocity and a change in the $\Afr$ measurement. This is not an issue for Gaussian or top-hat spectra because the effect will be the same on both sides of the profile as they only have one peak. Double-horn spectra typically have well-defined straight edges, so we would expect little change in the location of their measurement limits.  Second, if a target galaxy is not centered in the telescopes beam its response will not be spatially uniform over the \HI\, causing the emission to be detected favourably on one side of the profile. However, typical pointing offsets with the Arecibo radio telescope are 10-20 arcsec, and negligible with respect to its 3.5 arcmin beamsize. Third, there is the possibility of contaminant emission from neighboring galaxies. xGASS has been carefully assessed for confusion and we have discarded the spectra classified as confused.  We do not expect any of these three effects to be significant in our sample. Last, the spectral baseline subtraction is a source of uncertainty for measured \HI\ parameters, and might affect $\Afr$ estimates in either direction. However, this is common to any studies of global HI spectra and a significant issue only for low $\SN$ data.

\section{The rate of asymmetries in \lowercase{x}GASS} \label{sec:xGasym}
\begin{figure}
    \centering
    \includegraphics[width = 0.5\textwidth]{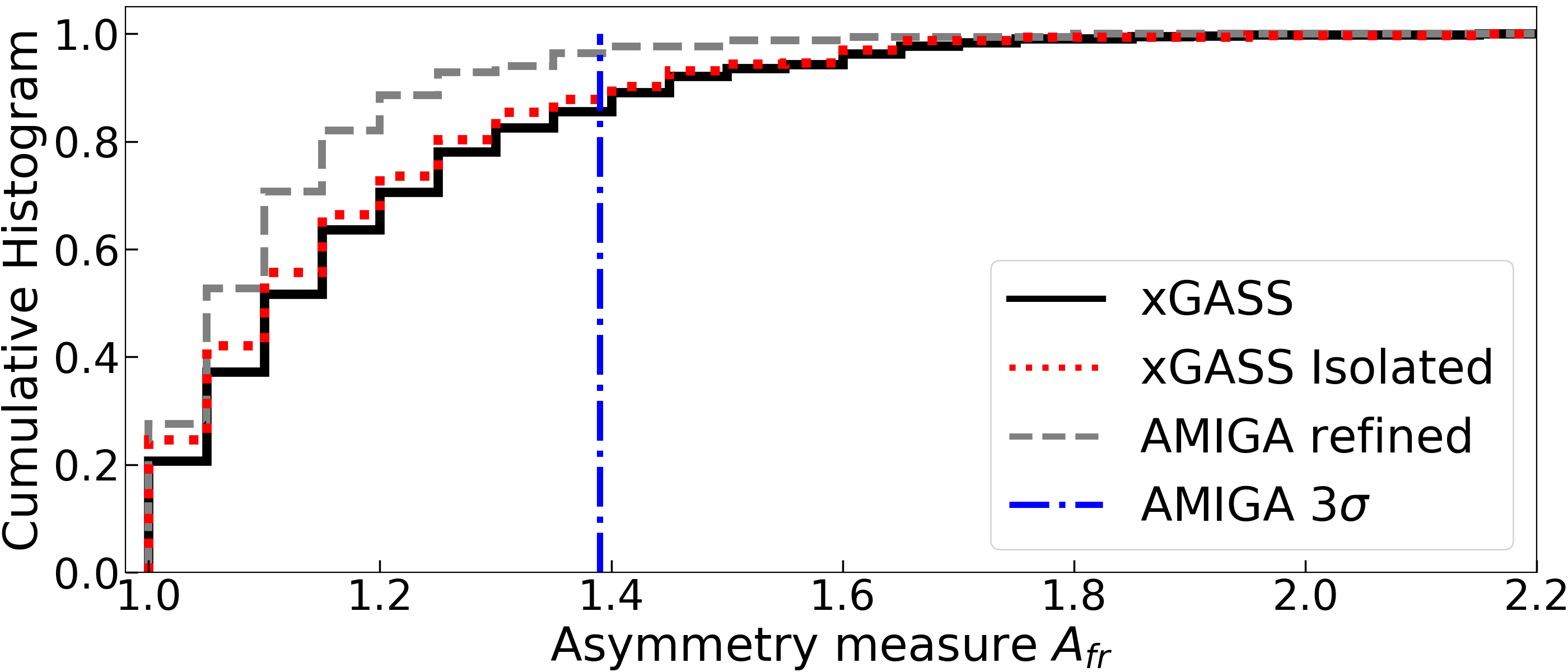}
\caption{xGASS asymmetry distribution: comparison between the cumulative $\Afr$ histogram for xGASS (black) and the  AMIGA refined sub-sample (grey, dashed). xGASS galaxies classified as isolated centrals are shown by the red dotted histogram, and the blue dot-dash vertical line is the AMIGA 3$\sigma$ asymmetry threshold.}
    \label{fig:xGA_compare}
\end{figure}

In Fig. \ref{fig:xGA_compare} we show the cumulative $\Afr$ distribution of  our xGASS sub-sample  in bins of $\Delta \Afr = 0.05$ as the solid black line. A spectrum with a perfectly symmetric profile has, by definition, $\Afr = 1$. \ct{haynes98} classified asymmetric spectra as $\Afr>1.05$, finding that 67\% of galaxies are asymmetric; and \ct{espada11} parameterised their $\Afr$ distribution using a half Gaussian with $\sigma=0.13$ finding 2\% of galaxies with  $\Afr > 3\sigma\ (1.39)$. As these studies used samples of isolated galaxies, we calculate the rate of asymmetries in xGASS using only the isolated central galaxies (the red dotted line in Fig. \ref{fig:xGA_compare}) and find an asymmetry rate of 76\% and 14\%, respectively. xGASS therefore shows a higher rate of asymmetry than these samples. 

Fig. \ref{fig:xGA_compare} also compares xGASS to the AMIGA refined sub-sample from \ct{espada11}, defined as spectra with an estimated $\Afr$ uncertainty of $<0.05$. The xGASS distribution has a lower cumulative fraction in all $\Afr$ bins compared to the \ct{espada11} distribution, confirming our result above that xGASS has a higher rate of asymmetries in global \HI\ spectra. Quantitatively, half of the xGASS galaxies have $\Afr>1.14$, whereas 31\% of the AMIGA galaxies show asymmetries greater than this level. This is unsurprising as xGASS contains galaxies in all environments and \HI\ asymmetries have been shown to be more frequent in denser environments \cp{angiras06,angiras07,scott18,bok19}, whereas the AMIGA galaxies are selected to be extremely isolated and should show the lowest rate of asymmetries.

\begin{figure}
    \centering
    \includegraphics[width = 0.5\textwidth]{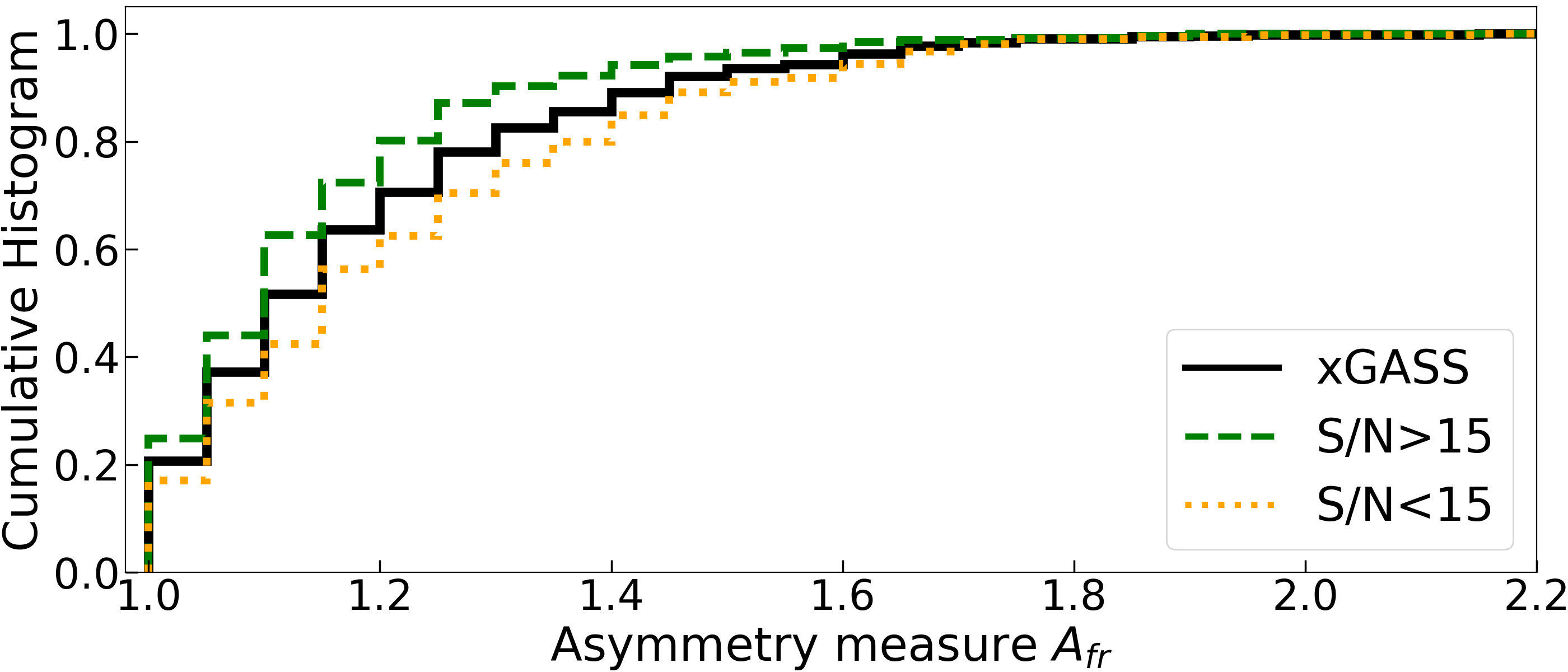}
    \caption{Cumulative $\Afr$ histograms for xGASS galaxies with $\SN>15$ and $\SN<15$, compared to whole xGASS sample.}
    \label{fig:Afr_SNdeff}
\end{figure}

In Fig. \ref{fig:Afr_SNdeff} we compare the cumulative $\Afr$ distributions for xGASS galaxies with $\SN>15$ and $\SN<15$. Clearly, spectra with lower $\SN$ have a higher rate of asymmetries compared to the higher $\SN$ spectra. This suggests that the fraction of galaxies classified as asymmetric may depend on the $\SN$ distribution of the sample. To properly assess the rate of asymmetries in a sample we must understand the effect of noise on the measurement of $\Afr$.  Uncertainties on individual $\Afr$ values have been estimated previously. \ct{espada11} combine their observational and measurement uncertainties, and find  typical $\sigma \Afr =  0.04$ with a maximum around 0.17.  \ct{bok19} estimated a maximum uncertainty of $<5\%$ at  $\SN=10$, their lower $\SN$ threshold, by repeatedly perturbing each channel in their spectra by a variate drawn from a Gaussian of width set by the observed $\rms$.

\section{The effect of noise on measured asymmetry} \label{sec:noisecorr}
\subsection{Mock \HI\ spectra} \label{subsec:mockspectra}
To understand the effects of measurement noise on our asymmetry measurements we generate mock global \HI\ spectra observations using a toy model. A total \HI\ mass ($\MHI$) is distributed in an infinitesimally thin disc according to an adopted radial surface density profile and projected with inclination $i$ onto a two dimensional sky-plane image. Each pixel is assigned an observed line-of-sight (LOS) velocity using a rotation curve (RC) template from \ct{catinella06} and corrected for projection effects using
\begin{equation}
    V_{\rm obs} = V(r)\cos(\phi)\sin(i),
\end{equation}
where $\phi$ is the azimuthal angle measured counter clockwise from the receding major axis. We use the \ct{catinella06} template RCs as they provide a good parameterisation of disc galaxies over a range of luminosities out to a maximum radius of 2 optical radii ($\Ropt$, defined as the radius enclosing 83\% of the integrated stellar light), which we adopt as the maximum extent of our model to avoid extrapolation. The template RCs adopt the Polyex model \cp{giovanelli02}
\begin{equation}
    V_{\rm PE}(r) = V_0\Big(1-e^{-r/R_{\rm PE}}\Big)\Big(1+\frac{\alpha r}{R_{\rm {PE}}}\Big),
\end{equation}
characterised by the amplitude $V_0$, exponential scale length $R_{\rm PE}$ and outer slope $\alpha$.
The radial \HI\ surface density profile is modelled as a constant surface density which transitions to an exponential decline with slope $\beta$ at transition radius $R_T$:
\begin{equation}
    M_\text{\HI}(r) = 
    \begin{cases} 
        1 & r <  R_T \\
        e^{- \beta (r - R_T)} & r \geq R_T.
    \end{cases}
\end{equation}
This is motivated by observations which show a central saturation of 8 $\SMpc$ interior to the optical radius, and decline exponentially at higher radii \cp{leroy08}. We adopt an exponential slope motivated by \ct{bigiel12} and \ct{wang14} who showed that the \HI-dominated outer parts of disc galaxies have an approximately universal slope of  $\beta=1.65/\Ropt$\footnote{Formally \ct{bigiel12} and \ct{wang14} scale their profiles by $R_{25}$, the radius of the 25 mag arcsec$^{-2}$ isophote, but for an exponential Freeman disc with central surface brightness of 21.6 mag arcsec$^{-2}$ it can be shown that $\Ropt \approx R_{25}$.}. The spectrum is observed in bins of rest-frame LOS velocity at resolution $\Vres$ and  the mass in each bin is converted to a spectral flux ($S_\nu$) for an adopted distance $d$ (and corresponding redshift $z_\text{\HI}$) by inverting 
\begin{equation}
    M_\text{\HI}[M_{\odot}] = \frac{2.356 \times 10^5}{(1+z_\text{\HI})} \; \Big(\frac{d}{\text{Mpc}}\Big)^2 \;  \int S_\nu \, d\text{v}\ \text{Jy}\ \kms.
\end{equation}
We also model a Gaussian spectrum with standard deviation $\sigma_{\rm G}$ and a top-hat spectrum of width $\omega_{\rm T}$ which are normalised to the same $\Sint$ as the \HI\ toy model spectrum. Measurement noise is modelled by perturbing each channel by a random variate drawn from $\mathcal{N}(0,\rmsin)$, and the spectrum is boxcar smoothed to velocity resolution $\Vsm$ using a boxcar of width $N_{\rm sm} = \Vsm/\Vres$. 

We generate noiseless, symmetric model spectra using the above model to create a  Gaussian spectrum with $\sigma_{\rm G} = 90\ \kms$, a narrow Gaussian with $\sigma_{\rm G} = 14\ \kms$, a top-hat spectrum with width $\omega_{\rm T} = 300\ \kms$, and a double-horn spectrum  using the following input parameters: 
\begin{itemize}
    \item $\MHI =10^{10}\ M_{\odot}$
    \item $d = 150$ Mpc 
    \item $i= 50^\circ$
    \item $R_T = R_{\rm opt}$
    \item $\beta = 1.65/R_{\rm opt}$
    \item $V_0 = 200\ \kms$
    \item $R_\text{PE} = 0.164\ R_\text{opt}$
    \item $\alpha  = 0.002\ R_\text{opt}$.
\end{itemize}
The double-horn and top-hat spectra are representative of galaxies with \HI\ distributions which extend to the flat part of the RC resulting in well defined, straight profile edges. The Gaussian profile represents the case when the \HI\ predominantly traces the rising part of the RC resulting in slowly rising edges, and the narrow Gaussian reflects the \HI\ spectrum of a near face-on galaxy where there is little LOS velocity information.

\begin{figure}
    \centering
    \includegraphics[width = 0.5 \textwidth]{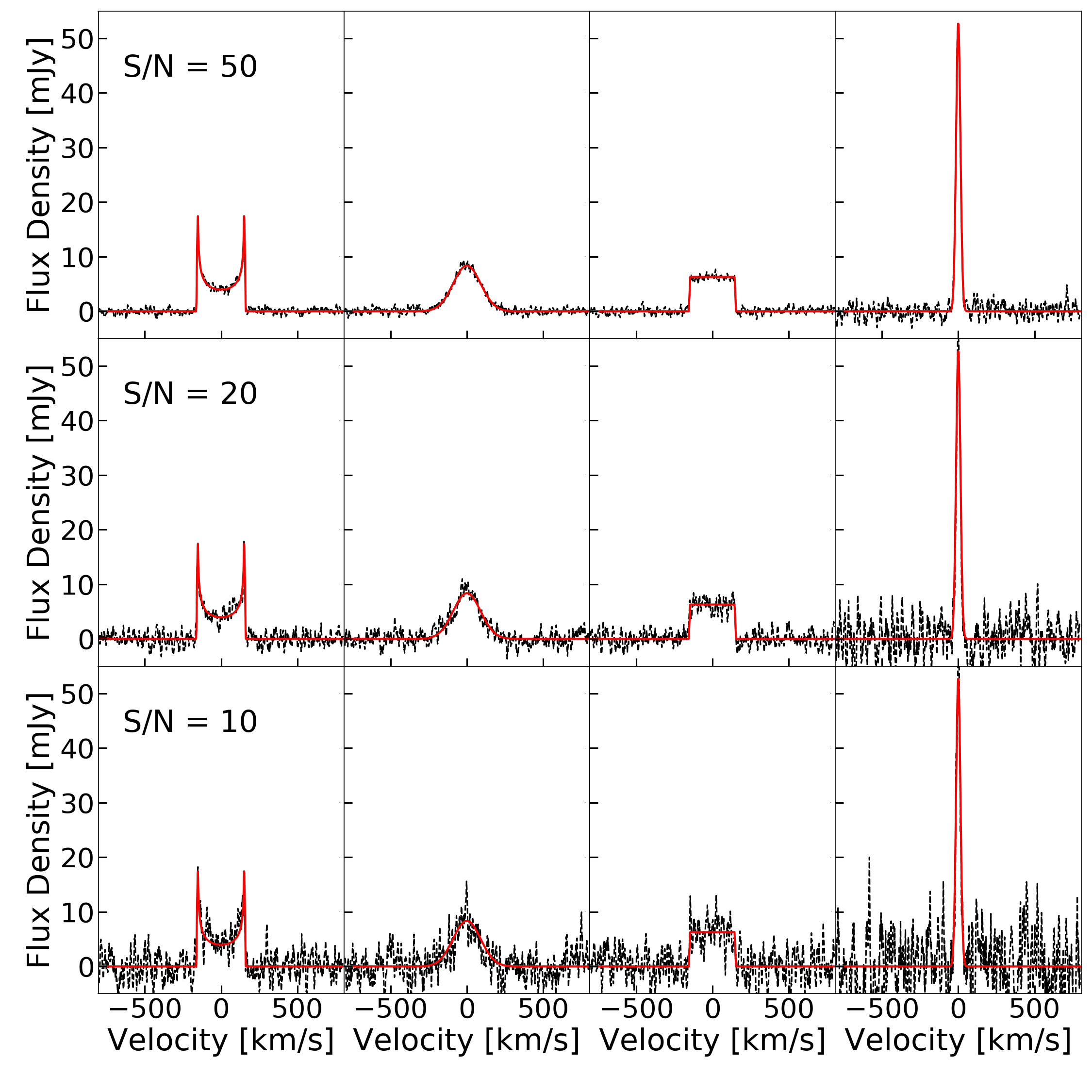}
    \caption{Mock observations of global \HI\ spectra for the double-horn (column 1), Gaussian (column 2), top-hat (column 3), and narrow Gaussian (column 4) models with $\SN = 50$ (top row), 20 (middle row) and 10 (bottom row). The noiseless model spectra are overlaid in red.}
    \label{fig:mockspectra}
\end{figure}

The model spectra are observed at $\Vres = 2\ \kms$ and smoothed to $\Vsm = 10\ \kms$. The limits of each template are defined in the same way as the observations and used to measure $\Sint$ and $\wtw$. The model spectra are used to create mock observations with desired $\SN$ by adding random noise with magnitude $\rmsin$ to each channel prior to smoothing. Knowing $\Sint$, $\wtw$, and $\Vsm$ of the model spectrum we can rearrange eq. \ref{eq:SN} to calculate $\rms$; the observed noise after smoothing. The required input noise $\rmsin$ is given by the noise reduction due to smoothing, $\rms = \rmsin / \sqrt{N_{\rm sm}}$, where $N_{\rm sm}$ is the width of the boxcar kernel. Sets of $10^4$ mock observations are created for $\SN \in [5,100]$ at steps of $\Delta\SN=1$ to properly sample a range covering our lower limit $\SN =7$ to very high quality spectra. We also generate the same sets of mock observations using the $\SN$  defined as the peak flux in the spectrum divided by $\rms$, $\SNPN$, as it is a common definition used in the literature \cp[e.g.][]{haynes98, espada11}. In Fig. \ref{fig:mockspectra} we show examples of mock double-horn, Gaussian, top-hat, and narrow Gaussian spectra with $\SN = 50,\ 20$ and $10$ with the model spectrum over-laid. We treat the model spectra as `fits' to the mock spectra and use the same measurement limits to   calculate $\Sint,\ \SN,\ \SNPN$ and $\Afr$ for all the mock observations. 

\subsection{The dependence of $\Afr$ on $\SN$} \label{subsec:SNafrcor}

\begin{figure}
    \centering
    \includegraphics[width = 0.5 \textwidth]{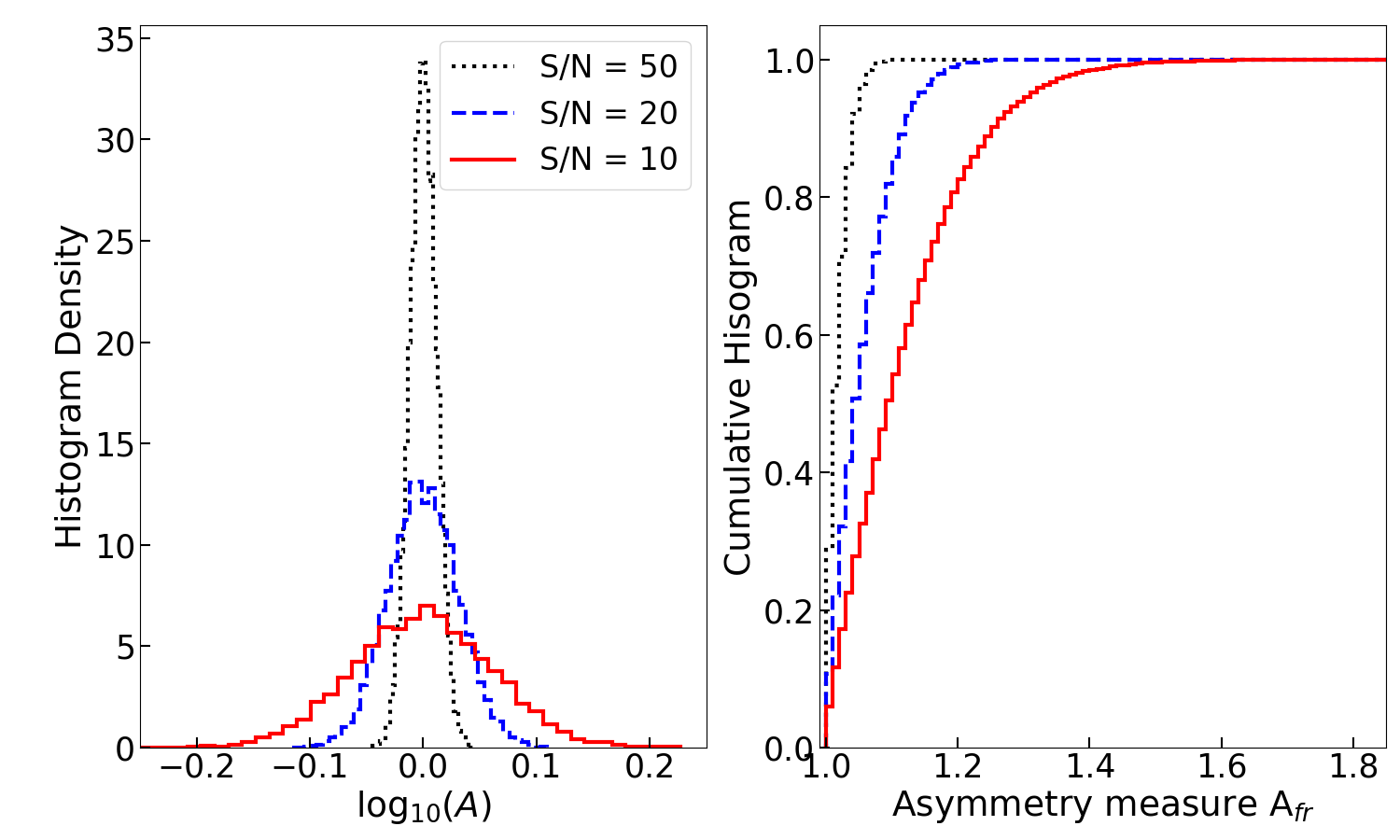}
    \caption{The effect of noise on asymmetry. Density normalised histogram of $\log_{10}(A)$ (left) and cumulative $\Afr$ distribution (right) for 10$^4$ mock observations of an intrinsically symmetric double horn spectrum with $\SN = 50$ (black, dotted), $\SN = 20$ (blue, dashed), and $\SN = 10$ (red, solid).}
    \label{fig:noise_asym}
\end{figure}

In the left panel of Fig. \ref{fig:noise_asym}, we show density-normalised histograms of \footnote{ As $A$ is a ratio quantity bounded over the range $[0,\infty)$, with a mean of $A=1$ for a symmetric spectrum, it obeys log-normal statistics. Thus, we show the distribution of $\log_{10}(A)$ such that any scatter is Gaussian.} $\log_{10}(A)$ for 10$^4$ mock observations of an intrinsically symmetric double-horn spectrum with input $\SN = 50,\ 20,$ and $10$.  While the average recovered value for each distribution is $\log_{10}(A) = 0$ (i.e. a perfectly symmetric spectrum) there is visible scatter about the mean, which increases with lower $\SN$. This increase in scatter is also clear in the right panel of Fig. \ref{fig:noise_asym}, where we show the cumulative $\Afr$ distributions of the same samples: i.e. the histograms reach a given cumulative fraction at higher $\Afr$ as $\SN$ decreases. This demonstrates that, while the distribution of recovered asymmetries is consistent with Gaussian noise, and the most probable $\Afr$ measurement is a symmetric spectrum, the uncertainty on an $\Afr$ measurement increases as $\SN$ decreases. In other words, a  perfectly symmetric spectrum will be measured to have $\Afr>1$ more frequently than it will be measured to have $\Afr=1$. We refer to this effect as ``noise-induced asymmetry". At $\SN=10$, $\sim$50\% of intrinsically symmetric spectra have noise-induced asymmetries greater than $\Afr = 1.1$, which is larger than typical $\Afr$ uncertainty estimates \cp[$\lesssim 10$\%;][]{espada11, bok19}. This demonstrates why individual $\Afr$ values cannot be directly compared: it is difficult to assess  the  degree  of  noise  contamination  in  an $\Afr$ measurement. Instead we must compare cumulative $\Afr$ distributions between samples and control for their different $\SN$. We present our method to compare $\Afr$ distributions in \S\ref{subsec:envasym}.

\begin{figure*}
    \centering
    \includegraphics[width = \textwidth]{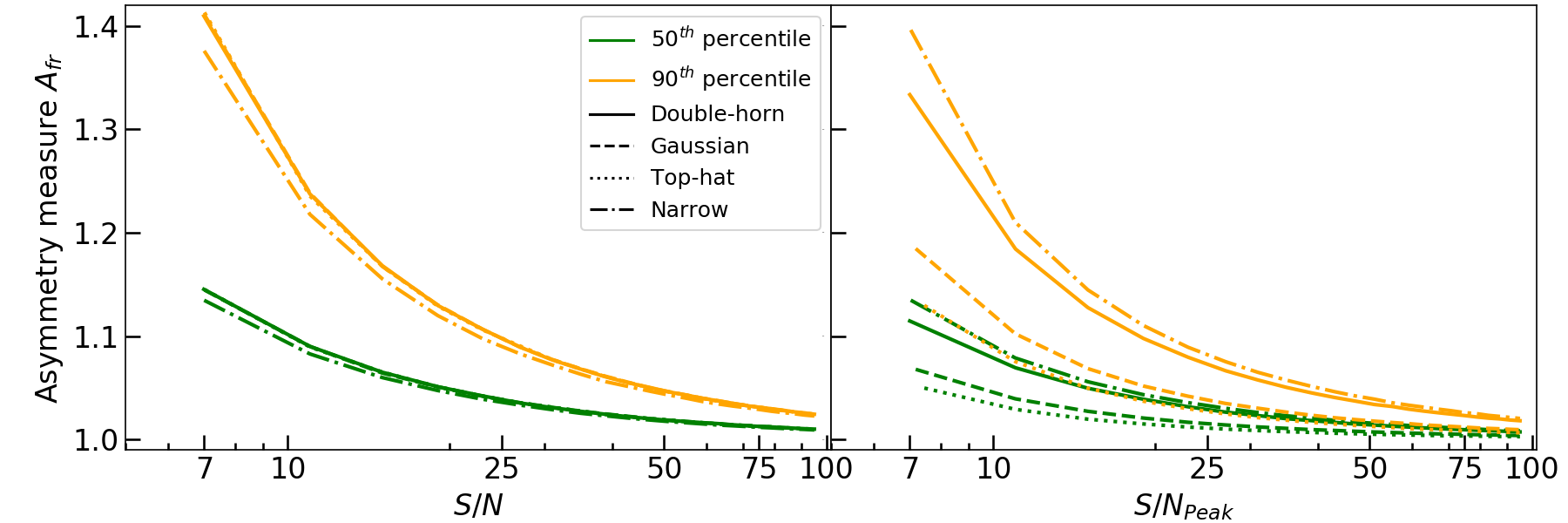}
    \caption{Asymmetry distribution as a function of $\SN$ for different \HI\ profile shapes. The 50$^{\rm th}$ (green) and 90$^{\rm th}$ (orange) percentiles of the asymmetry distribution are shown as a function of the average $\SN$ in each bin for $\SN$ (left) and $\SNPN$ (right). The solid line corresponds to the double-horn model, the dashed to the Gaussian,  the dotted to the top-hat, and the dot-dashed to the narrow Gaussian.  }
    \label{fig:shapecompare}
\end{figure*}

To investigate how this noise-induced asymmetry behaves for different spectrum shapes and $\SN$ measures we place our mock spectra in bins of $\Delta (\SN) = \Delta (\SNPN) = 4$ and quantify the width of the $\Afr$ distribution in each bin by calculating its 50$^{\rm th}$ and 90$^{\rm th}$ percentiles\footnote{As $A$ is a log-normal quantity $\Afr$ is a half log-normal distribution, which is not meaningfully described by Gaussian statistics, so we use percentiles.}. In Fig. \ref{fig:shapecompare} we show these percentiles as a function of the mean $\SN$ in each bin. The same trend is shown as in Fig. \ref{fig:noise_asym}: the width of the $\Afr$ distribution increases as $\SN$ decreases, and this trend is also visible with $\SNPN$.

The most striking feature of  Fig. \ref{fig:shapecompare} is that the magnitude of noise-induced asymmetries depends  on the choice of $\SN$ measurement and the shape of a spectrum. The double-horn, Gaussian and top-hat percentiles are almost identical as a function of $\SN$, and the narrow Gaussian percentile shows the same behaviour but sits slightly below the others; indicating that very narrow spectra are not as easily affected by noise. In the lowest $\SN$ bin of the 90$^{\rm th}$ percentile this difference is $\sim 9\%$. Our inclusion of the narrow Gaussian spectrum demonstrates that the similarity between the double-horn, Gaussian, and top-hat percentiles is not simply a result of them having similar $\rms$. The narrow Gaussian requires a higher $\rms$ to have the same $\SN$ at fixed $\Sint$, as is seen in Fig. \ref{fig:mockspectra}, but despite this its percentiles show minimal difference compared to the other models.  All the percentiles have similar behaviour as a function of $\SN$ because it is a robust measure regardless of profile shape. The right panel of Fig. \ref{fig:shapecompare} shows that the percentiles are not similar at fixed $\SNPN$. A double-horn spectrum with similar $\Sint$ and $\wtw$ as a Gaussian has a higher peak flux (see Fig. \ref{fig:mockspectra}) and requires a higher $\rms$ to have the same $\SNPN$. This higher $\rms$ can cause greater noise-induced asymmetries in the double-horn spectrum compared to the Gaussian, hence the double-horn percentile sits above the Gaussian one in the right panel of Fig. \ref{fig:shapecompare}.  

\begin{figure*}
    \centering
    \includegraphics[width = \textwidth]{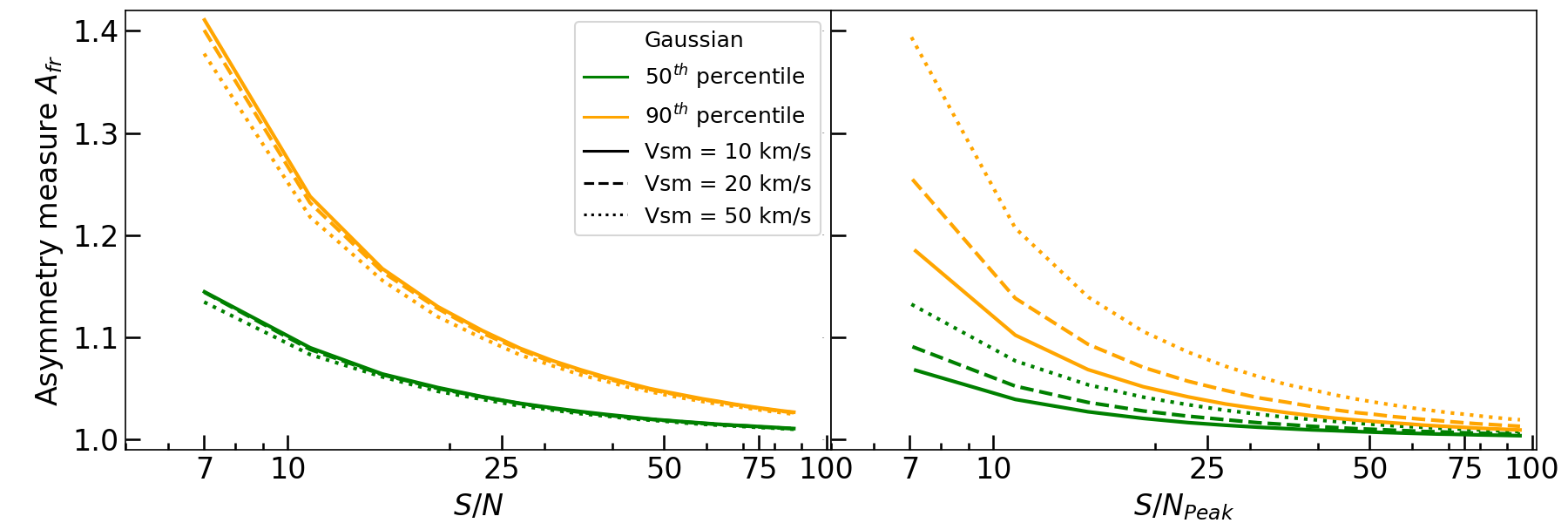}
    \caption{Asymmetry distribution as a function of noise for a Gaussian \HI\ profile with different degrees of smoothing. The 50$^{\rm th}$ (green) and 90$^{\rm th}$ (orange) percentiles of the asymmetry distribution for a Gaussian spectrum are shown as a function of the average $\SN$ in each bin for $\SN$ (left) and $\SNPN$ (right). The solid line corresponds to a final smoothed velocity resolution of $\Vsm = 10\ \kms$, the dashed to $\Vsm = 20\ \kms$ and  dotted to $\Vsm = 50\ \kms$. }
    \label{fig:smoothcompare}
\end{figure*}

In Fig. \ref{fig:smoothcompare} we show the 50$^{\rm th}$ and 90$^{\rm th}$ percentiles of the $\Afr$ distribution for a Gaussian spectrum  after boxcar smoothing to final velocity resolutions of  10, 20 and 50 $\kms$ as a function of $\SN$ and $\SNPN$. $\SN$ shows little variation between the different degrees of smoothing as it is, by definition, invariant to smoothing; whereas at fixed $\SNPN$ spectra with higher smoothing have a wider $\Afr$ distribution. This is due to the behaviour of $\SNPN$ as boxcar smoothing conserves flux, and thus $\Afr$. Smoothing decreases $\rms$ by a factor of $\sqrt{N_{\rm sm}}$ and increases profile width (which decreases the peak) by $\sim 0.5 \Vsm$ \cp{catinella13}, though the $\rms$ reduction is the dominant effect. A $\Vsm$ of 10 $\kms$ at $\Vres = 2\ \kms$ decreases the peak of the narrow Gaussian by 9\% and the $\rms$ by a factor of $\sqrt{5}$, increasing $\SNPN$ by a factor of $\sim 2$. At fixed $\SNPN$ spectra with higher smoothing can have noise-induced asymmetries drawn from the distribution corresponding to their pre-smoothed $\rms$ (i.e. $\rmsin$).

Fig. \ref{fig:shapecompare} and \ref{fig:smoothcompare} indicate that we must exercise caution when interpreting and comparing the $\Afr$ distributions of samples with different distributions of $\SN$, and if $\SNPN$ is used profile shapes and smoothing add additional uncertainty.  $\SN$ is clearly the preferred choice if we wish to control for noise-induced asymmetry, so we only consider it from here.
The shape of the $\SN$ percentiles are well described by the functional form 
\begin{equation} \label{eq:percentile}
    P(\SN) = 1\ +\ \frac{1}{a \, (\SN - b)},
\end{equation}
where $a$ and $b$ are functions of the desired percentile. We provide the values of $a$ and $b$ for fits to the 40$^{\rm th}$ to 95$^{\rm th}$ percentiles of a double-horn spectrum in Table. \ref{tab:Pfits}, though they are applicable to any profile shape.

\begin{table}
 \caption{Coefficients $a$ and $b$ from fits to the 40$^{\rm th}$ to 95$^{\rm th}$ percentiles for a double-horn spectrum \label{tab:Pfits}}
 \begin{tabular}{p{2.25cm}|p{2.25cm}|p{2.25cm}}
 \hline
Percentile &  a & b  \\ \hline \hline
P40 & 1.363 & 0.375 \\
P45 & 1.196 & 0.411 \\
P50 & 1.063 & 0.513 \\
P55 & 0.951 & 0.602 \\
P60 & 0.851 & 0.656\\
P65 & 0.768 & 0.760 \\
P70 & 0.693 & 0.864 \\
P75 & 0.624 & 0.945 \\
P80 &  0.560 & 1.049 \\
P85 & 0.499 & 1.179 \\
P90 & 0.438 & 1.415\\ 
P95 & 0.372 & 1.850\\
\hline
 \end{tabular}
\end{table}

\section{Robust quantification of symmetric and asymmetric \HI\ spectra} \label{sec:asympop}
\begin{figure}
    \centering
    \includegraphics[width= 0.5\textwidth]{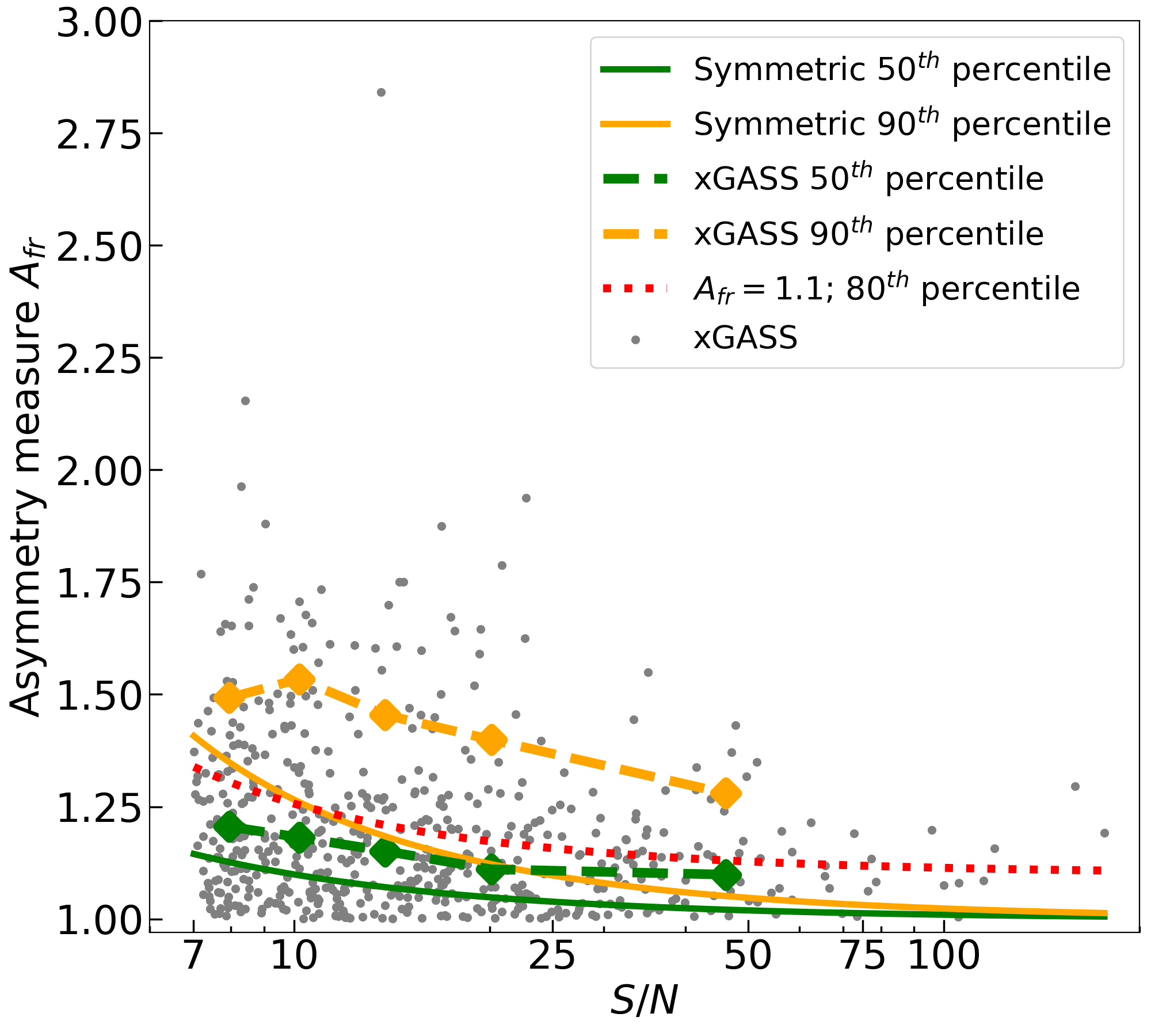}
    \caption{The $\SN-\Afr$ parameter space for xGASS.  Grey background points are individual xGASS galaxies and the thicker dashed lines are the 50$^{\rm th}$ (green) and 90$^{\rm th}$ (orange) percentiles of the $\Afr$ distribution in bins of $\SN$, the bin centres are shown as diamonds.  The thinner solid lines are 50$^{\rm th}$ (green) and 90$^{\rm th}$ (orange) percentiles for a double-horn spectrum from. \S\ref{subsec:SNafrcor},and the thin red line is the 80$^{\rm th}$ percentile of a spectrum with intrinsic $\Afr=1.1$.}
    \label{fig:SNasym}
\end{figure}

With an understanding of the effect of noise on $\Afr$, we can identify spectra which are confidently asymmetric and place more meaningful constraints on the rate of \HI\ profile asymmetry in xGASS. In Fig. \ref{fig:SNasym} we plot the $\Afr$ of the xGASS galaxies as a function of their $\SN$ (grey points), with the 50$^{\rm th}$ and 90$^{\rm th}$ percentiles for a double-horn spectrum from \S\ref{subsec:SNafrcor} overlaid (thin solid green and orange lines). We also overlay the same percentiles calculated for the xGASS galaxies in bins with edges $\SN=$ 7, 8.84, 11.6, 16.5, 25.8, \& 177, defined to have the same number of galaxies in each bin (112), as dashed lines centered at the mean $\SN$ in each bin shown as diamonds. The xGASS percentiles are above the model percentiles across all $\SN$ values, so  \emph{xGASS exhibits a higher rate of global \HI\ asymmetries than what can be attributed to measurement noise.} The typical \HI\ spectrum of a galaxy detected in our deep xGASS observations is therefore not symmetric. 

One major consequence of the effect of noise on $\Afr$ is that our $\Afr$ distribution has become a $\SN - \Afr$ parameter space: the definition of an asymmetric spectrum must take into account $\SN$. This makes the percentiles calculated in \S\ref{subsec:SNafrcor} an ideal tool for defining populations as there is 90\% confidence that a measured $\Afr$ $\geq$ P90 (given its $\SN$) is due to an intrinsic asymmetry in the spectrum. To define a population consistent with being symmetric we select galaxies with $\Afr\ \leq$ P50, as 50\% of initially perfectly symmetric galaxies will have $\Afr$ below this threshold. At low $\SN$ there will be some contamination from galaxies with small asymmetries, but the deviation from symmetry shown by these galaxies is small compared to the scatter introduced by noise and we do not expect them to affect our results.

\begin{figure}
    \centering
    \includegraphics[width= 0.5\textwidth]{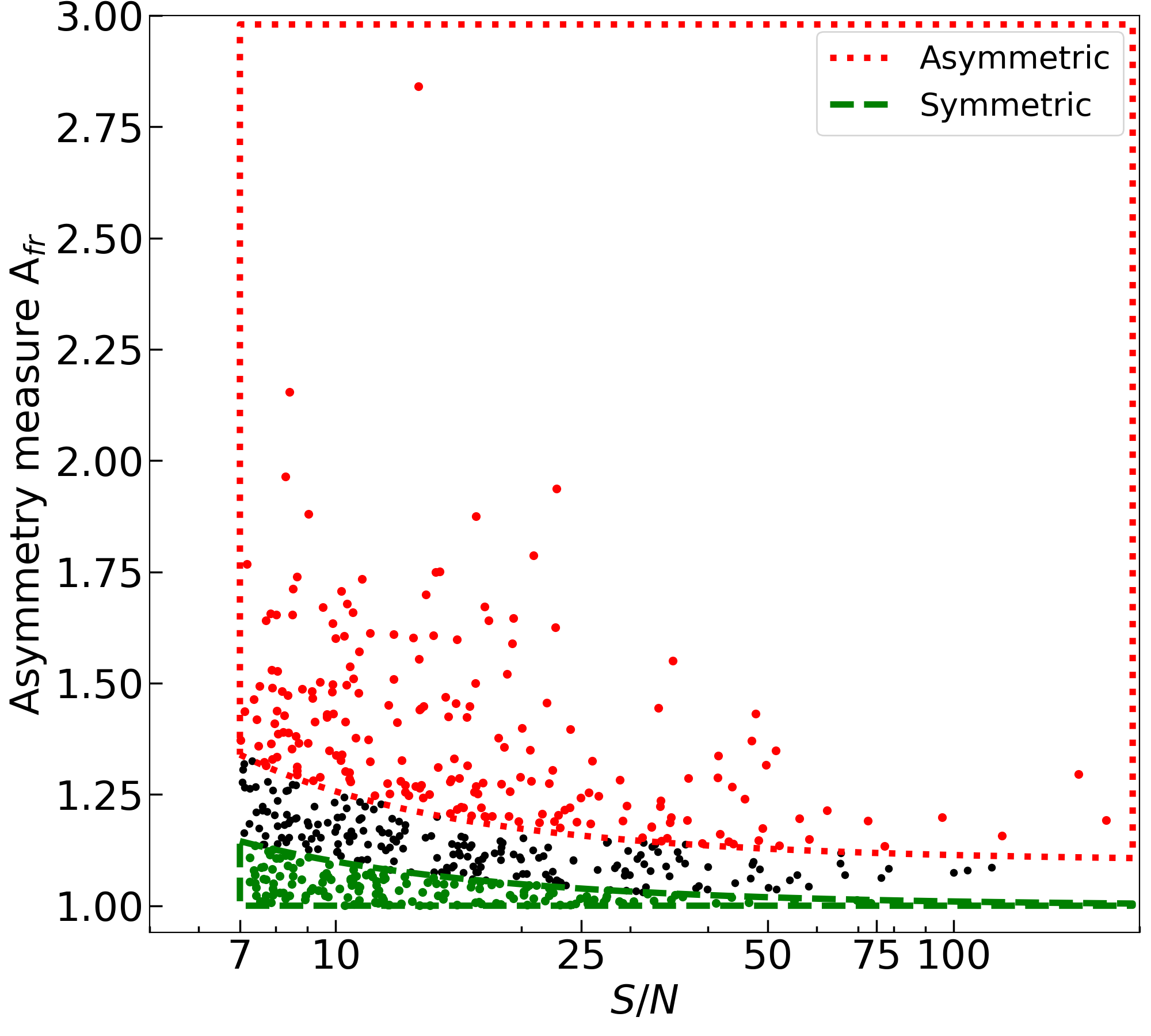}
    \caption{Definition of symmetric (green, dashed line) and asymmetric (red, dotted line) galaxies in the $\SN-\Afr$ parameter space.}
    \label{fig:popsel}
\end{figure}

To define our asymmetric population we cannot use the percentiles from \S\ref{subsec:SNafrcor} as they have been computed using symmetric spectra and their use would include spectra with intrinsic asymmetry close to $\Afr=1$. To avoid this contamination, and because we are not interested in the few galaxies with highly reliable (high $\SN$) but tiny levels of asymmetry, we define $\Afr=1.1$ as our threshold for asymmetry. This choice is somewhat arbitrary, but is similar to the values used by \ct{haynes98} and the 1$\sigma$ (= 0.13) of the \ct{espada11} $\Afr$ distribution. Using the same procedure outlined in  \S\ref{sec:noisecorr} we generate a double-horn spectrum with intrinsic asymmetry of $\Afr=1.1$ and model the $\SN-\Afr$ parameter space. The 80$^{\rm th}$ percentile of the $\Afr$ distribution as a function of $\SN$ is then used to define the asymmetric population, i.e. intrinsic $\Afr\geq 1.1$ with 80\% confidence. This percentile is shown in Fig. \ref{fig:SNasym} as a red dotted line. The percentage of asymmetric galaxies in the same bins used to calculate the xGASS percentiles in Fig. \ref{fig:SNasym} is 33.6\%, 33.6\%, 37.5\%, 38.4\%, and 38\%, which corresponds to overall asymmetry rate of 37\%. The rate of asymmetry is relatively uniform as a function of $\SN$, which gives us confidence that we are properly correcting for the effects of noise. The asymmetric and symmetric populations contain 207 and 155 galaxies respectively, and their corresponding selection boxes are are shown in Fig. \ref{fig:popsel}.

\begin{figure*}
    \centering
    \includegraphics[width=\textwidth]{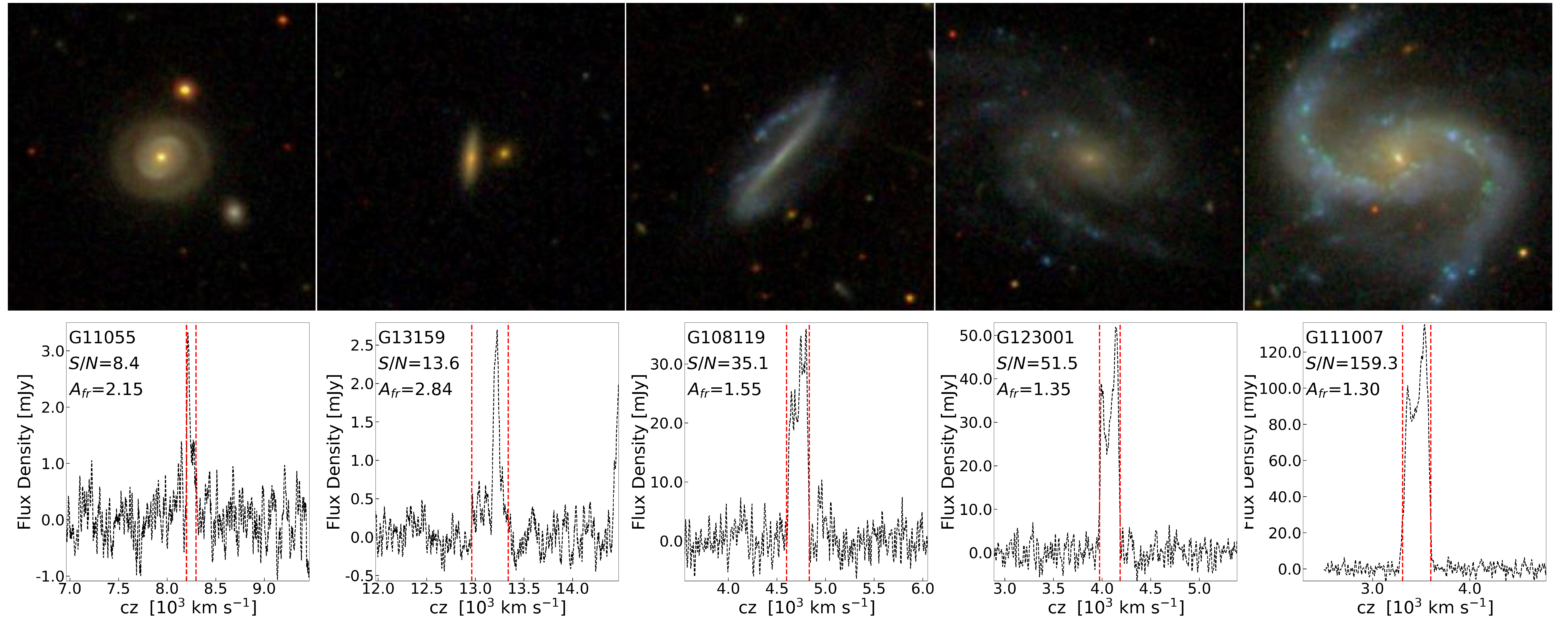}
    \caption{SDSS postage stamps (1.5 arcmin square) and global \HI\ spectra for the five most asymmetric galaxies selected in equal width $\log_{10}\SN$ bins. The limits used for measuring each spectrum are shown as red dashed lines, and the measured $\SN$ and $\Afr$ are shown in the top left corner of each spectrum. }
    \label{fig:spec_opt_asym.png}
\end{figure*}

\begin{figure*}
    \centering
    \includegraphics[width=\textwidth]{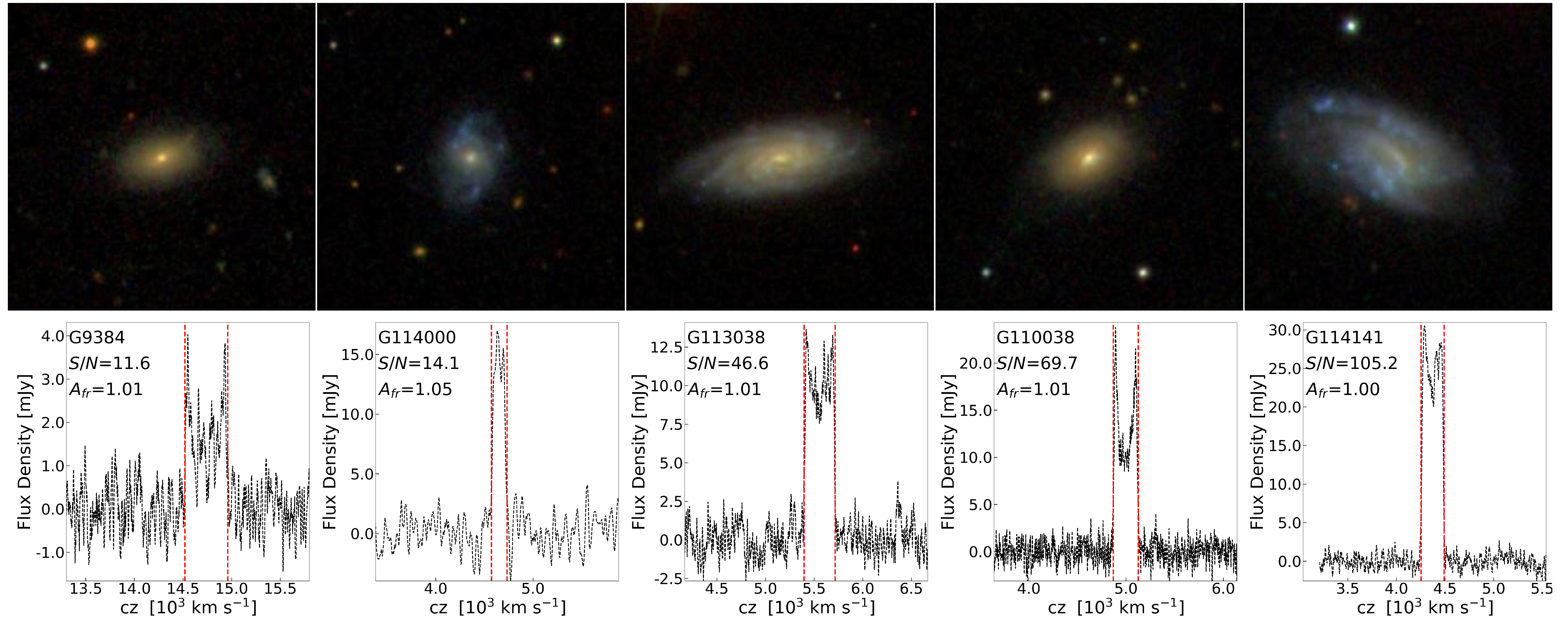}
    \caption{The same as Fig. \ref{fig:spec_opt_sym.png} but for five symmetric galaxies in the same $\log_{10}\SN$ bins.}
    \label{fig:spec_opt_sym.png}
\end{figure*}

 In Figs. \ref{fig:spec_opt_asym.png} and \ref{fig:spec_opt_sym.png}  we show the five most asymmetric spectra and five symmetric spectra selected in equal width $\log_{10}\SN$ bins alongside  SDSS optical images of their host galaxies. The diversity in optical properties displayed by both samples provides no immediate distinction between \HI\ symmetric and asymmetric galaxies. Both samples contain blue star-forming and red passive galaxies, galaxies showing signs of optical disturbances, as well as ones that appear optically undisturbed. This is consistent with previous studies which observe little correlation between \HI\ profile asymmetry and optical asymmetries or morphology \cp[e.g.][]{haynes98,matthews98,kornreich00,wilcots04,espada11,bok19}. The most asymmetric galaxy in the sample, G13159,  is the galaxy second from the left in Fig. \ref{fig:spec_opt_asym.png} and is also the strongest outlier at the top in Fig. \ref{fig:SNasym}. The xGASS source notes point out two small blue objects without optical redshifts that are visible to the East of the galaxy, but aside from these there are no likely companions within the beam and it is not classified as confused.

\section{Relationship between asymmetry and galaxy properties} \label{sec:asymenv}

\subsection{\HI\ properties of the asymmetric population}
With lack of a strong qualitative correlation between global \HI\ asymmetry and optical morphology, we utilise the deep \HI\ observations of xGASS to investigate the \HI\ content of the asymmetric population. To compare the average \HI\ content of an asymmetric galaxy to the typical content for symmetric galaxies of similar stellar mass, we match each asymmetric galaxy to all symmetric galaxies within 0.15 dex in stellar mass and 0.05 dex in $\SN$. If less than five galaxies are matched we iteratively expand each of the bins by 0.01 dex up to limits of 0.25 dex in stellar mass and 0.15 dex in $\SN$, and we exclude galaxies which exceed these tolerances. The $\SN$ matching is included as low $\SN$ galaxies are biased toward asymmetry, and $\SN$ correlates positively with gas fraction at fixed stellar mass which could lead us to infer that asymmetric galaxies are gas-poor compared to symmetric ones. The definition of our populations restricts the number of symmetric galaxies at high $\SN$ such that the highest $\SN$ of a matched asymmetric galaxy is $\SN = 33.6$. Quantitatively, 166 (80\%) of the asymmetric galaxies were successfully matched to five or more symmetric ones.

\begin{figure}
    \centering
    \includegraphics[width= 0.5\textwidth]{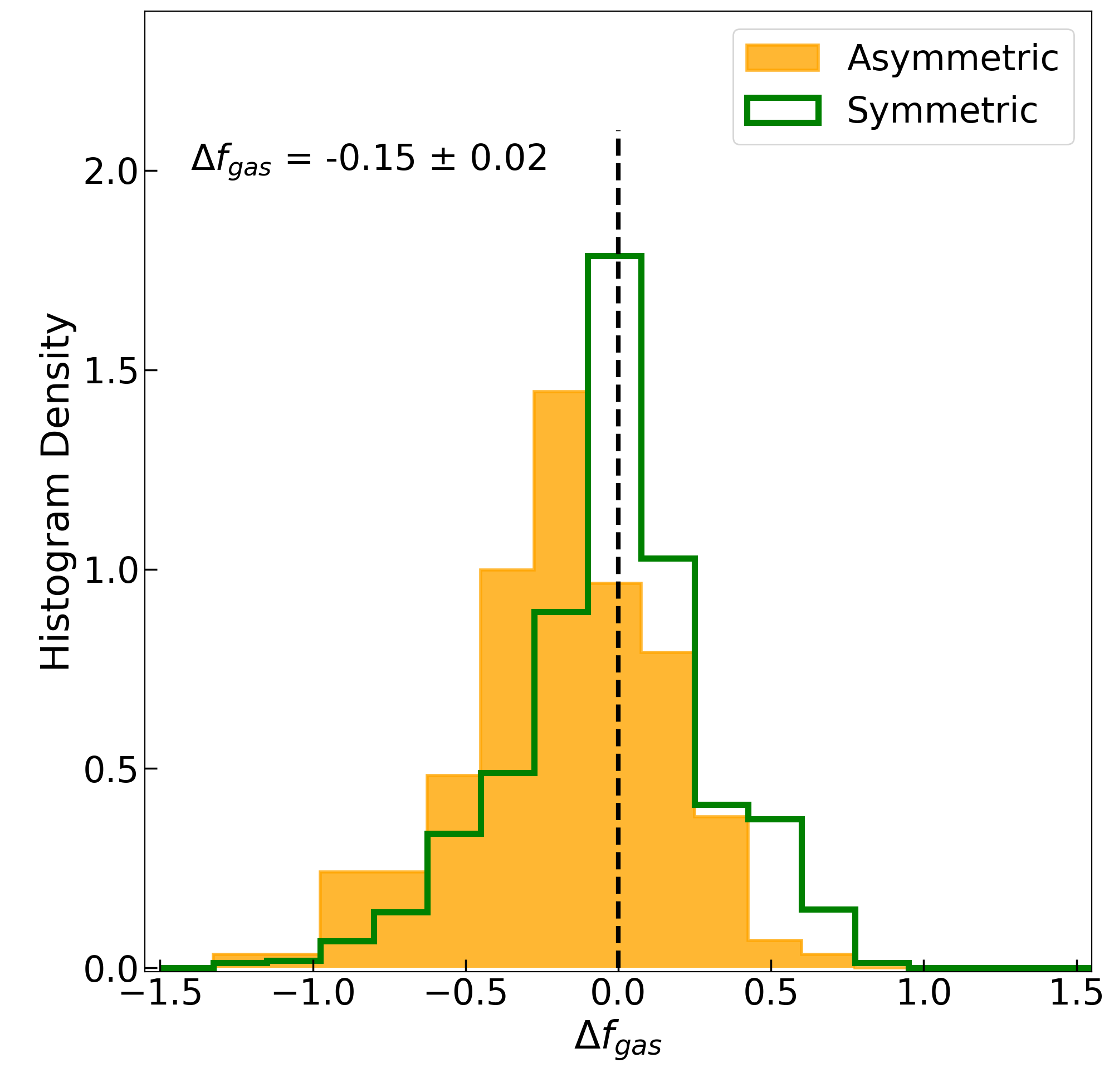}
    \caption{Density normalised histograms of \HI\ gas fraction offsets controlled by stellar mass and $\SN$. The asymmetric galaxies are shown by the  orange, filled histogram and the symmetric galaxies by a  green, open histogram. The median $\dfg$ for the asymmetric galaxies is given in the top left corner  and  the vertical dashed line corresponds to the median offset of the symmetric galaxies, $\dfg=0$.}
    \label{fig:deltafgas}
\end{figure}

We compute the gas fraction offset \cp[$\dfg$,][]{ellison18} as the difference between the logarithmic gas fraction  of an asymmetric galaxy and the median logarithmic gas fraction of its matched symmetric galaxies. The  $\dfg$ distribution for the asymmetric galaxies is shown in Fig. \ref{fig:deltafgas} along with the distribution for the matched symmetric galaxies which, by definition, is centered at $\dfg = 0$. Asymmetric galaxies show a systematic offset toward negative $\dfg$ values with a median $\dfg = -0.15 \pm 0.02$, where the uncertainty on the median is calculated from bootstrapping the asymmetric population $10^4$ times. This demonstrates that, on average,  \emph{asymmetric galaxies contain 29\% less \HI\ than their stellar mass and $\SN$ matched, symmetric counterparts}. This result is robust to our choice of percentiles used to define the populations. If we choose P70 or P90 (for a spectrum with input $\Afr=1.1$) to select the asymmetric population, the resulting median gas fraction offsets are $\dfg = -0.14 \pm 0.02$ and $\dfg = -0.17 \pm 0.03$ respectively. Similarly, if we select the symmetric population using P40 or P60 (for a symmetric input spectrum) the median gas fraction offsets are $\dfg = -0.12 \pm 0.03$ and $\dfg = -0.16 \pm 0.03$ respectively. 

\begin{figure}
    \centering
    \includegraphics[width= 0.5\textwidth]{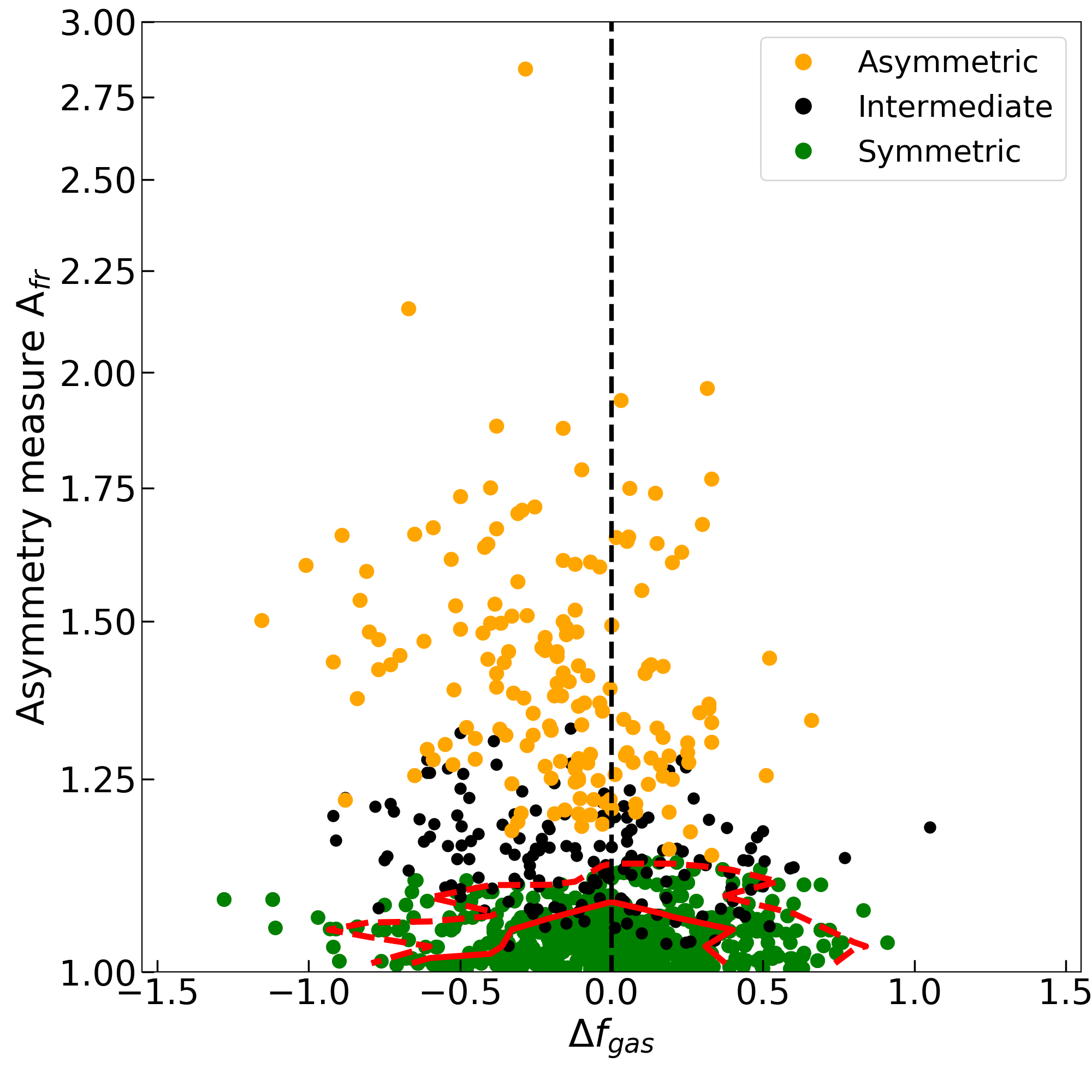}
    \caption{Comparison between $\dfg$ and $\Afr$.  Asymmetric galaxies are shown in orange, symmetric galaxies  (orange), and the intermediate population in black. The red solid and dashed contours define the region which encapsulate 68\% and 95\% of the symmetric galaxies, respectively.}
    \label{fig:dfg_Afr}
\end{figure}

To investigate which galaxies are driving this trend we compare $\dfg$ and $\Afr$ for the asymmetric, symmetric, and the intermediate (black points in Fig \ref{fig:popsel}) populations in Fig. \ref{fig:dfg_Afr}.  The red contours denote the region that contain  68\% (solid) and 95\% (dashed) of the symmetric galaxies.  We see that asymmetric galaxies, as in Fig. \ref{fig:deltafgas}, are offset toward negative $\dfg$ values; however there is also a trend where galaxies with higher $\Afr$ are more likely, on average, to have more negative $\dfg$ values. This is also visible in the intermediate population of galaxies. If we divide the population in two based on their asymmetry we find the more asymmetric half contains, on average, 40\% less \HI\ than the symmetric half. This trend is not likely a consequence of noise-induced asymmetry, as we have $\SN$ matched our spectra in the $\dfg$ calculation.  More asymmetric galaxies are more gas-poor. 

Our asymmetric galaxies are therefore tracing a population where \HI\ is being disturbed and removed. One interpretation of this is that environmental processes, such as ram pressure or tidal stripping \cp{gunn72,kenney04,stevens17}, might be the dominant driver of \HI\ asymmetries in xGASS.  Quantitatively 25\% and 15\% of the matched asymmetric and symmetric samples (respectively) are satellites, indicating a stronger contribution of galaxies potentially undergoing environmental effects to the asymmetric sample.

\subsection{Environment as a driver of asymmetry} \label{subsec:envasym}

Motivated by our observation of lower gas fractions and higher satellite fractions in the asymmetric sample, we compare the asymmetry distributions of galaxies in xGASS in different environments. As mentioned in \S\ref{subsec:SNafrcor} to fairly compare two $\Afr$ distributions we must sample them from the same $\SN$ distribution. Given two samples (e.g. satellite galaxies, central galaxies), we compute their density normalised $\SN$ histograms in bins of 0.05 dex between $\SN=7$ and 50, and treat galaxies with $\SN>50$ as being in the same bin as noise effects are negligible at higher $\SN$. We define the `common' $\SN$ distribution between the two samples as the minimum density in each respective bin of the two corresponding $\SN$ histograms. The ratio of the common histogram to a sample's  histogram gives the  fraction of galaxies in each bin needed to match it.  Galaxies are selected by generating a uniform random variate in the range [0,1] and only keeping those with variates lower than the fraction in their corresponding $\SN$ bin. We repeat this process 10$^4$ times, calculate the cumulative $\Afr$ distribution in bins of $\Delta \Afr = 0.05$ for the selected galaxies in each sample, and record the mean and standard deviation for each bin. 

\begin{figure}
    \centering
    \includegraphics[width= 0.5 \textwidth]{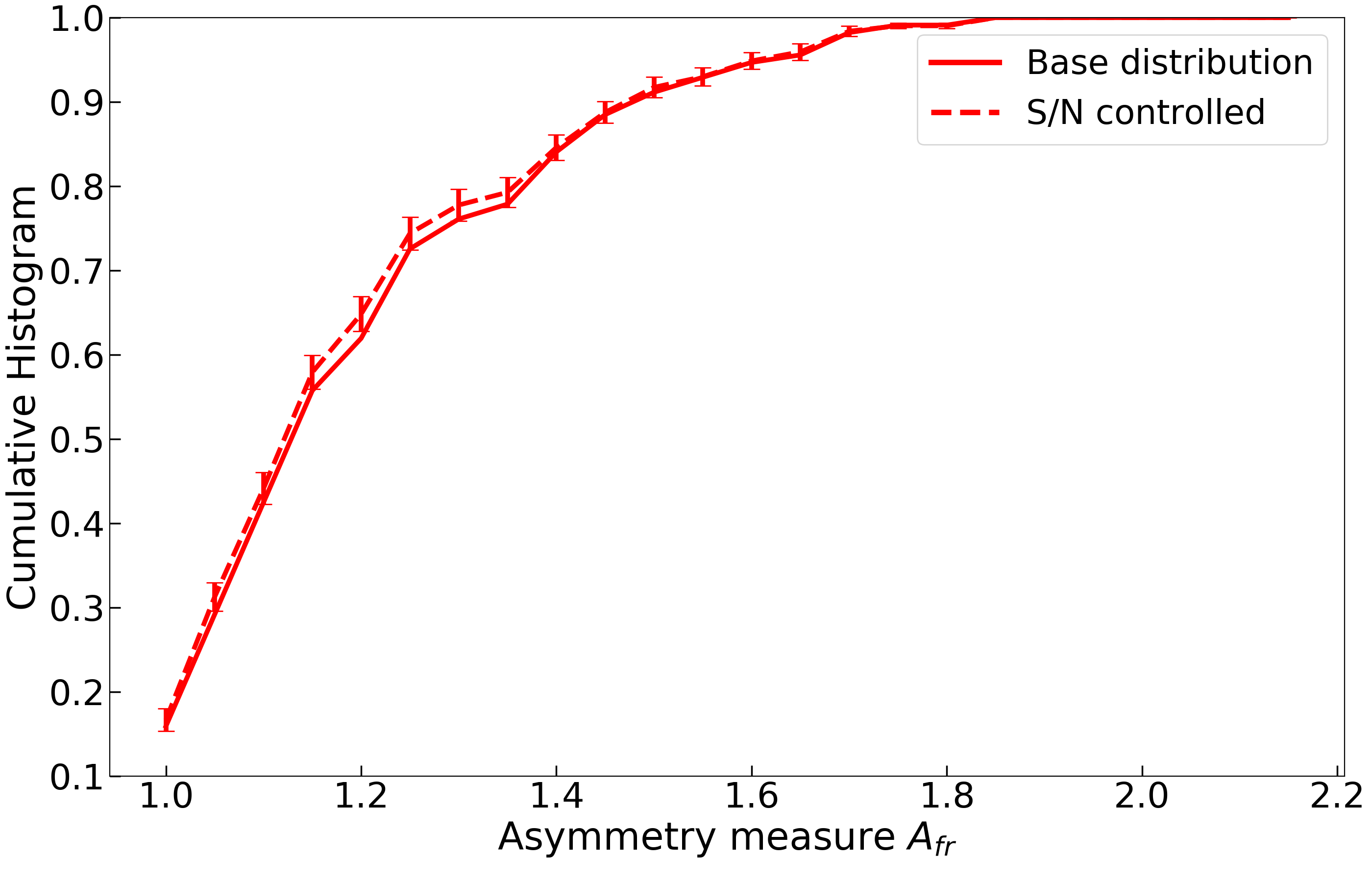}
    \caption{The cumulative $\Afr$ distribution of satellite galaxies (solid) compared to the average distribution (dashed) from $10^4$ iterations of sampling from the $\SN$ control histogram defined between the satellites and centrals. Error bars on the average histogram are the 1$\sigma$ standard deviation of the $10^4$ distributions in each bin.}
    \label{fig:SNsamp}
\end{figure}

In Fig. \ref{fig:SNsamp} we compare the cumulative $\Afr$ distribution of the satellite galaxies to their mean cumulative $\Afr$ distribution with 1$\sigma$ error bars, after controlling the central and satellite galaxies to their common $\SN$ histogram. The mean cumulative distribution is equal or marginally higher in each $\Afr$ bin compared to the uncontrolled distribution, so it is a slightly more symmetric sample of galaxies in comparison. This is because the satellite sample has relatively more low $\SN$ spectra in comparison to the centrals, so the sampling process acts to reduce their contribution and subsequently the contribution from spectra with high $\Afr$ due to noise. The uncontrolled distribution is within the error bars of the mean cumulative, so the $\SN$ sampling does not significantly change the shape of the cumulative $\Afr$ distribution (in this case). 

In addition to the uncertainty introduced by our $\SN$ sampling, we assess the magnitude of the uncertainty originating from our sample using the delete-a-group jackknife (DAGJK) algorithm \cp{kott01}.  The algorithm calculates the weighted difference between parameter estimates with a random unique subset removed, and the mean parameter estimate from removing $N$ unique subsets. The DAGJK standard deviation estimator is 
\begin{equation}
    \sigma(x) \ =\ \sqrt{\rm{var}(x)} \ =\ \sqrt{ \frac{N-1}{N} \sum_{i =1}^{N} \Big(x_i - \bar{x} \Big)^2},
\end{equation}
where $x_i$ is the parameter estimate with the $i^{\rm th}$ subset removed and $\bar{x}$ is the mean of the $N$ $x_i$ estimates. Removing galaxies randomly from a sample will not necessarily conserve the $\SN$ distribution, which we have carefully controlled to ensure a fair comparison of $\Afr$ distributions. Instead of deleting a fraction of the sample, which is the common DAGJK method,  we delete this  fraction from each bin of the sample's $\SN$ histogram. This effectively conserves the $\SN$ distribution between each DAGJK  iteration.   We delete a random 20\% of galaxies from a sample in each iteration, so the number of iterations $N$ is five. The DAGJK uncertainty is calculated for each bin in the cumulative $\Afr$ distribution for each of the $10^4$ $\SN$ sampling iterations described above. The median uncertainty in each $\Afr$ bin is typically 2-3 times larger than the uncertainty estimated from the $\SN$ sampling, so the uncertainty is dominated by the variation within the sample. 

\begin{figure}
    \centering
    \includegraphics[width= 0.5 \textwidth]{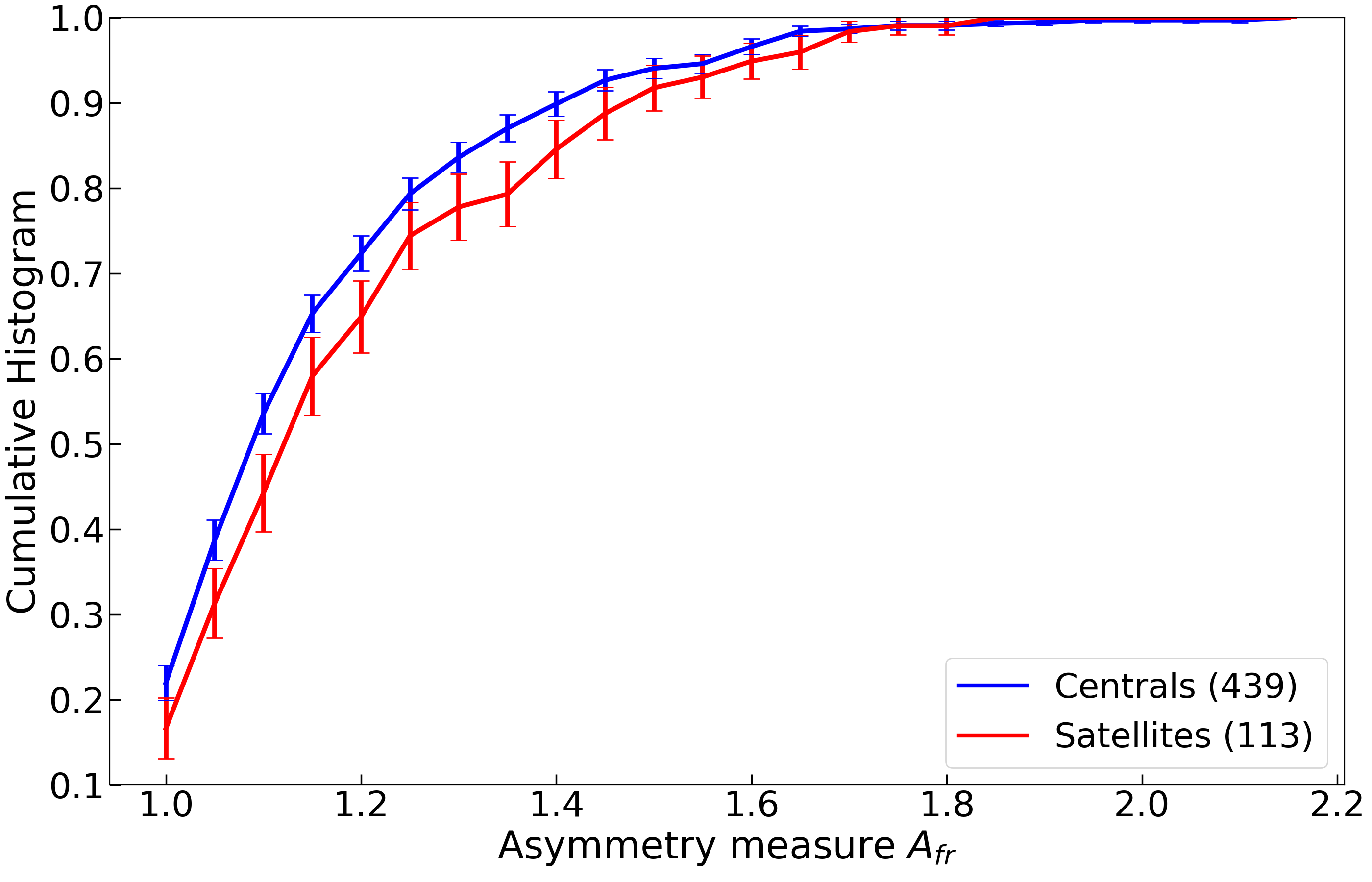}
    \caption{Average cumulative $\Afr$ distributions of central (blue) and satellite (red) galaxies controlled to the same $\SN$ distribution. Error bars correspond to the median DAGJK uncertainty in each bin, and the numbers in the legend correspond to the number of galaxies in each sample.  }
    \label{fig:Afrhistsatcent}
\end{figure}

In Fig. \ref{fig:Afrhistsatcent} we show the mean cumulative $\Afr$ distributions for the satellite and central galaxies with median 1$\sigma$ DAGJK uncertainty from $10^4$ $\SN$ sampling iterations as error bars. The cumulative distribution of the satellite galaxies sits below the centrals until they converge around a cumulative fraction of 0.9 at $\Afr=1.5$. The clear separation of the two distributions over the majority of this range confirms that satellite galaxies exhibit a higher rate of global \HI\ asymmetries than central galaxies, at least in our xGASS sample. This is consistent with \ct{angiras06, angiras07, scott18} and \ct{bok19}, who show that \HI\ asymmetries are more frequent in denser environments. Thus our work provides the first clear evidence that, as a population, satellite galaxies have more asymmetric global \HI\ spectra than centrals. 

Although satellite galaxies have typically lower mass compared to centrals, this is not reflected by our xGASS sub-sample as shown in Fig. \ref{fig:sample}. To investigate whether stellar mass may influence the asymmetry of a galaxy, we performed the same analysis after splitting our sample into galaxies with masses above and below $M_{\star}=10^{10}\ M_{\sun}$, and found no difference between the cumulative $\Afr$ histograms. Environment is therefore the main driver of global \HI\ asymmetries in xGASS.

\begin{figure}
    \centering
    \includegraphics[width= 0.5 \textwidth]{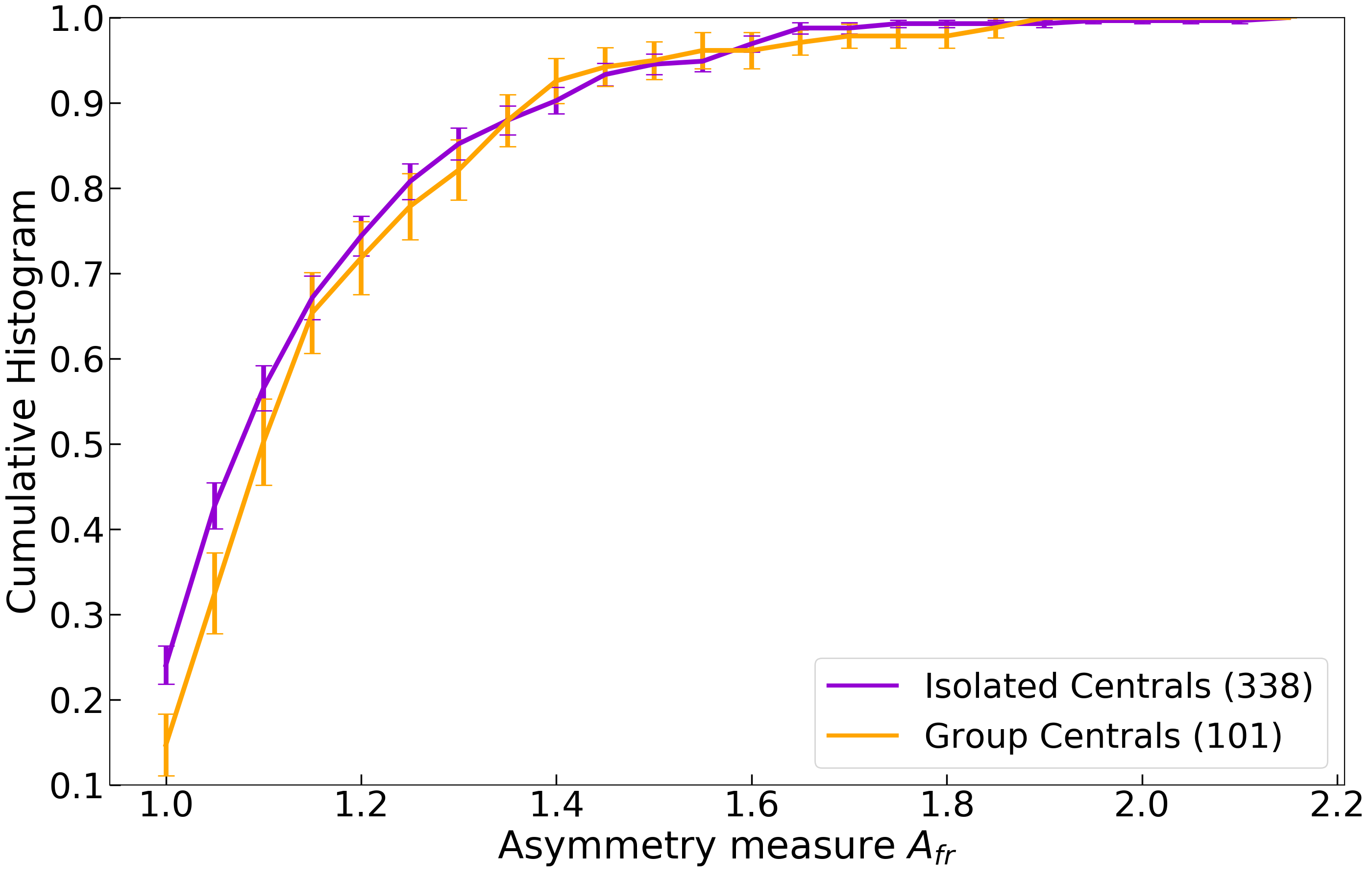}
    \caption{The same as Fig. \ref{fig:Afrhistsatcent} but for isolated central (purple) and group central (orange) galaxies.} 
    \label{fig:Afrhistisogrp}
\end{figure}

We have made the distinction between satellite and central galaxies, but it is also interesting to compare the $\Afr$ distributions of isolated and group central galaxies. Fig. \ref{fig:Afrhistisogrp} shows the mean cumulative $\Afr$ distributions with median DAGJK 1$\sigma$ error bars for isolated and group central galaxies. The two distributions show no marked differences, except in the first two $\Afr$ bins, where there is a clear difference in the fraction of galaxies with $\Afr<1.1$.  This suggests that group centrals show a higher rate of small disturbances in their \HI\ compared to isolated centrals, and is consistent with asymmetries being more frequent in denser environments. 

\section{Discussion} \label{sec:discuss}

\subsection{Implications for global \HI\ asymmetry studies}
Using flat thresholds regardless of $\SN$, previous studies have estimated that at least 50\% of galaxies have asymmetric global \HI\ spectra \cp[e.g.][]{haynes98,matthews98}. In this work we found that the distribution of measured asymmetries is affected by the $\SN$ of the observations, such that a flat threshold would also select spectra with smaller intrinsic asymmetry. This suggests that the rates of global \HI\ asymmetry found by previous studies are likely upper limits and dependent on their sample's $\SN$ distribution. However, it does not detract the statement made by \ct{richter94} that ``asymmetries in disc galaxies may be the rule, rather than the exception"; we still observe the distribution of global \HI\ asymmetries to be wider than what can be attributed to noise (Fig. \ref{fig:SNasym}), at least in a stellar mass-selected sample such as xGASS.

It also demonstrates that the definition of a reference $\Afr$ distribution is difficult as extremely high $\SN$ observations would be needed to reach the true intrinsic distribution. Making comparisons to this distribution would also require the same high $\SN$ observations, which would reduce sample sizes and the statistical power gained by using global \HI\ spectra would be reduced. We have shown in this work that this is not necessary. It is possible to compare the $\Afr$ distributions of galaxy populations and make statements about their gas reservoirs; but it must be done in a relative sense, and taking into account profile $\SN$. 

This means that the \ct{espada11} $\Afr$ distribution remains a powerful quantification of the asymmetries due to secular evolution in isolated disc galaxies;  all that is required to make comparisons to their dataset are $\SN$ measurements of their spectra.

\subsection{Implications for galaxy evolution}
In this work we have found that asymmetric galaxies preferentially have less \HI\ at fixed stellar mass, and that satellite galaxies have more asymmetric global \HI\ spectra than centrals. This suggests that asymmetric galaxies are preferentially gas-poor \emph{and} satellites, implying that the physical mechanism for making a satellite galaxy gas-poor is also making the profile asymmetric.

This is consistent with the observed asymmetries in \HI\ deficient \cp{haynes84} galaxies undergoing ram pressure stripping in clusters \cp[e.g.][]{kenney04,vollmer04,abramson11}. Environmental suppression of \HI\ has been observed in satellite galaxies residing in $\sim10^{13.5}\ M_{\odot}$ mass haloes by \ct{brown17}, and \ct{catinella13} showed in the GASS sample that galaxies in haloes of mass $10^{13}-10^{14}\ M_{\odot}$ have at least 0.4 dex less \HI\ than those in lower density environments at fixed stellar mass. These halo masses are typical of the group environment, and are dominated by satellites at all stellar masses in both GASS, and xGASS. Using resolved \HI\ asymmetries in the Eridanus group \ct{angiras06} found that late-type galaxies, which are typically more gas-rich, are less asymmetric than early-types. While they suggest that tidal interactions are the main driver of the asymmetries and \HI\ deficiency in their galaxies, \ct{catinella13} and \ct{brown17} suggest that hydro-dynamical processes are the likely cause of the \HI\ suppression in their samples. This highlights some of the remaining uncertainties regarding which environmental process is the dominant driver of asymmetries.

Our observed correlation between suppressed \HI\ content and asymmetry appears inconsistent with some previous studies. Compared to the group environment, isolated late-type galaxies show  higher rates of global \HI\ asymmetry \cp{matthews98,haynes98} and stronger, more frequent asymmetry in their stellar components \cp{zaritsky97,bournaud05}.  \ct{ellison18} showed that post-merger galaxies are preferentially gas-rich at fixed stellar mass, and \ct{bok19} showed that galaxies in close pairs have more asymmetric global \HI\ spectra than isolated galaxies. Although many of these studies do not make direct comparisons of gas content at fixed stellar mass, these systems (mergers, gas-rich/late-type isolated galaxies) are preferentially gas-rich compared to the average xGASS galaxy, and therefore not representative as a population. In our Fig. \ref{fig:deltafgas} we see a tail of gas-rich, asymmetric galaxies which may correspond to these systems; but they are not the typical asymmetric galaxy in this sample.

Consistent with previous studies, we see that the typical \HI\ spectrum for a central galaxy is not symmetric \cp[e.g.][]{espada11}. Interactions with satellite galaxies may partially explain why we observe group centrals to be more asymmetric than isolated centrals, but it cannot explain our observation of asymmetric isolated centrals. This returns us to the long standing issue that asymmetries must be frequently excited or long-lived otherwise they would be washed out within a few dynamical timescales. Cosmological gas accretion is often suggested as a driver as observations show it to be irregular, in the form of cloud complexes, filaments, or gas-rich minor mergers \cp[see the review by][]{sancisi08}. \ct{portas11} and \ct{ramirez18} suggested their spatially resolved \HI\ asymmetries were caused by minor mergers, and the simulations by \ct{bournaud05} showed that filamentry accretion can replicate observed global \HI\ asymmetries which could also explain the high rates observed in late-type isolated galaxies \cp{matthews98}.

Alternatively, \HI\ asymmetries may trace disturbances in the gravitational potential of galaxies. Spatial asymmetries in the old stellar population in galaxies are common \cp{zaritsky97,bournaud05, angiras06, zaritsky13} and their amplitude typically increase with radius until it flattens, as traced by the \HI\ \cp[e.g][]{angiras06,angiras07}, suggesting it is a long lived, global perturbation \cp{vaneymeren11a,vaneymeren11b}. These studies suggest that asymmetries trace disturbances in the dark matter haloes of galaxies, or offsets between the centre of the stellar disk and its dark matter halo, which are excited by tidal interactions. The drivers of asymmetry in the gas-rich regime remain an open question, which must be addressed with larger samples and hydrodynamical simulations \cp[e.g.][]{elbadry18}.

\section{Conclusions} \label{sec:concl}
In this work, we presented an analysis of the asymmetries, measured as the ratio of the flux between the two halves of global \HI\ spectra, for 562 galaxies in the xGASS sample. Using these data, and a comprehensive analysis of how noise affects the measurement of  asymmetry, we investigated the \HI\ properties of asymmetric galaxies and inferred what the driving mechanisms are likely to be. Our main results are as follows:
\begin{itemize}
    \item The measurement of asymmetry depends strongly on the signal-to-noise ($\SN$) of a spectrum, and this correlation must be accounted for when comparing the asymmetry distributions of populations or when defining spectra as asymmetric.
    \item The typical global \HI\ spectrum in a stellar mass-selected sample such as xGASS is more asymmetric than what can be attributed to noise. 37\% of xGASS galaxies detected with $\SN\geq7$ show asymmetry greater than 10\% at an 80\% confidence level. Consistent with previous studies, we see no obvious correlation between \HI\ asymmetry and optical galaxy morphology.
    \item \HI\ asymmetric galaxies, defined as $\Afr>1.1$ with 80\% confidence, contain 29\% less \HI\ gas compared to their stellar mass and $\SN$ matched, symmetric counterparts.
    \item Satellite galaxies, as a population, show a higher rate of global \HI\ asymmetries compared to centrals, and group central galaxies show a higher rate of small asymmetries compared to isolated centrals. We conclude that environment is the main driver of asymmetry in xGASS.
\end{itemize}

A common observation made in previous studies is the lack of correlation between global \HI\ asymmetry and galaxy properties such as morphology, star formation rate, or optical asymmetries \cp{haynes98, matthews98,kornreich00,espada11,bok19}. In this work we have provided the means to confidently select asymmetric global \HI\ spectra, and a methodology to compare the asymmetry distributions of different galaxy populations. This opens the door to revisiting the correlations between \HI\ asymmetry and galaxy properties, and raises new questions such as the connection between the  global \HI\ profile and asymmetries in the resolved gas kinematics from integral field unit studies \cp[e.g.][]{bloom18} and spatially-resolved \HI\ observations \cp[e.g.][]{giese16}. In particular the bulge-disc decomposition of xGASS by \ct{cook19} presents a unique opportunity to search for, or confirm the lack thereof, a connection between global \HI\ asymmetry and the optical morphology and asymmetry of galaxies. These are essential components to building a more complete understanding of the gas reservoirs of galaxies.

\section*{Acknowledgements}
We thank the anonymous referee for their useful comments which improved this paper. ABW acknowledges the support of an Australian Government Research Training Program (RTP) Scholarship throughout the course of this work. LC is the recipient of an Australian Research Council Future Fellowship (FT180100066) funded by the Australian Government. Parts of this research were supported by the Australian Research Council Centre of Excellence for All Sky Astrophysics in 3 Dimensions (ASTRO 3D), through project number CE170100013.



\bibliographystyle{mnras}
\bibliography{bibfile} 



\appendix


\bsp	
\label{lastpage}
\end{document}